\documentclass[sigconf]{acmart}
\AtBeginDocument{%
  }
\setcopyright{acmlicensed}
\copyrightyear{2025}
\acmYear{2025}
\acmDOI{XXXXXXX.XXXXXXX}
\acmConference[MSR 2026]{23rd International Conference on Mining Software Repositories}{April 2026}{Rio De Janeiro, Brazil}
\acmISBN{978-1-4503-XXXX-X/2018/06}

\usepackage{fontawesome5}
\usepackage{xspace}
\usepackage[shortcuts]{extdash}
\usepackage[frozencache,cachedir=minted-cache]{minted}
\usepackage{ifthen}
\usepackage{enumitem}
\usepackage{tabularx}
\usepackage{colortbl}
\usepackage{multirow}
\usepackage{hhline}
\usepackage{xcolor}
\usepackage{url}
\usepackage{graphicx}
\usepackage{textcomp}
\usepackage{hyperref}
\usepackage{makecell}
\usepackage[most]{tcolorbox}
\usepackage{amsthm}
\usepackage[]{mdframed}

\begin{document}
\title{A Match Made in Heaven? AI-driven Matching of Vulnerabilities and Security Unit Tests}

%ACM
\author{Emanuele Iannone}
\orcid{0000-0001-7489-9969}
\affiliation{%
  \institution{Hamburg University of Technology}
  \city{Hamburg}
  \country{Germany}
}
\email{emanuele.iannone@tuhh.de}
\author{Quang-Cuong Bui}
\orcid{0000-0001-6072-9213}
\affiliation{%
  \institution{Hamburg University of Technology}
  \city{Hamburg}
  \country{Germany}
}
\email{cuong.bui@tuhh.de}
\author{Riccardo Scandariato}
\orcid{0000-0003-3591-7671}
\affiliation{%
  \institution{Hamburg University of Technology}
  \city{Hamburg}
  \country{Germany}
}
\email{riccardo.scandariato@tuhh.de}

\newboolean{showcomments}
\setboolean{showcomments}{false}
\newboolean{peerreview}
\setboolean{peerreview}{false}

\definecolor{gray75}{gray}{.25}
\definecolor{gray50}{gray}{.5}
\definecolor{gray40}{gray}{.6}
\definecolor{gray30}{gray}{.7}
\definecolor{gray25}{gray}{.75}
\definecolor{gray20}{gray}{.8}
\definecolor{gray15}{gray}{.85}
\definecolor{gray10}{gray}{.9}
\definecolor{gray05}{gray}{.95}
\definecolor{redbg}{HTML}{F2968F}
\definecolor{greenbg}{HTML}{CDE4AE}
\definecolor{ghdiffredbg}{HTML}{ffccce}
\definecolor{ghdiffgreenbg}{HTML}{abefbc}
\definecolor{goalcolor}{HTML}{fffff2}
\definecolor{rqboxcolor}{HTML}{f2f2ff}
\definecolor{rqanswercolor}{HTML}{faf9f5}
\definecolor{takeawaycolor}{HTML}{f2fff2}
\definecolor{darkgreen}{HTML}{009B55}

\newcommand{\goal}[1]{ %
	\medskip %
	\noindent\fcolorbox{black}{goalcolor}{%
		\parbox{0.97\linewidth}{% 
			\textbf{\faBullseye\ Our Goal.} #1 %
		}%
	}%
	\medskip %
}%

\mdfdefinestyle{boxstyle1}{%
    linecolor=black,
    linewidth=3pt,%
    leftmargin=0cm,rightmargin=0cm,%
    roundcorner=5pt,
    topline=false,bottomline=false,rightline=false,%
    backgroundcolor=gray05
}
\mdfdefinestyle{boxstyle2}{%
    linecolor=purple,linewidth=2pt,%
    leftmargin=0cm,rightmargin=0cm,%
    roundcorner=5pt,
    topline=true,bottomline=true,rightline=false,leftline=false,%
    backgroundcolor=gray05
}
\mdfdefinestyle{boxstyle3}{%
    linecolor=red,linewidth=0pt,%
    leftmargin=0cm,rightmargin=0cm,%
    roundcorner=5pt,
    topline=true,bottomline=true,rightline=false,leftline=false,%
    backgroundcolor=purple!05
}
\mdfdefinestyle{boxstyle4}{%
    linecolor=black,linewidth=2pt,%
    leftmargin=0cm,rightmargin=0cm,%
    roundcorner=5pt,
    topline=true,bottomline=true,rightline=false,leftline=false,%
    backgroundcolor=black!10,
    frametitlebackgroundcolor=black
}
\mdfdefinestyle{boxstyle5}{%
    linecolor=darkgreen,
    linewidth=3pt,%
    leftmargin=0cm,rightmargin=0cm,%
    roundcorner=5pt,
    topline=false,bottomline=false,rightline=true,leftline=true,%
    backgroundcolor=darkgreen!10
}
\newcommand{\prompt}[2]
{
    \medskip
    \begin{mdframed}[style=boxstyle4]
    {\underline{\textbf{#1}}\\#2}
    \end{mdframed}
    \smallskip
}
\newcommand{\rqbox}[2]
{
    \smallskip
    \begin{mdframed}[style=boxstyle1]
    {\textbf{\faSearch\ RQ$_{#1}$.} \textit{#2}}
    \end{mdframed}
    %\smallskip
}
\newcommand{\rqanswer}[2]
{
    \smallskip
    \begin{mdframed}[style=boxstyle5]
    {\textbf{\faHandPointRight\ Answer to RQ$_{#1}$.} #2}
    \end{mdframed}
}

\newcommand{\furtheranswer}[1]{%
	\medskip%
	\noindent\fcolorbox{black}{rqanswercolor}{%
		\parbox{0.97\linewidth}{% 
			\textbf{\faHandPointRight[regular]\ Further Analysis Summary.} #1%
		}%
	}%
	\medskip%
}%

\newcommand{\takeaway}[1]{%
	\medskip%
	\noindent\fcolorbox{black}{takeawaycolor}{%
		\parbox{0.97\linewidth}{% 
			\textbf{\faEnvelopeOpenText\ Takeaway Message.} #1 %
		}%
	}%
	\medskip%
}%

\newcommand{\nb}[3]{
    \fbox{\bfseries\sffamily\scriptsize\color{#3}#1}
	{\sf\small$\blacktriangleright${\color{#3}\textit{\textbf{#2}}}$\blacktriangleleft$}
}

\newcommand\rqOne{\textbf{RQ$_1$}\xspace}
\newcommand\rqTwo{\textbf{RQ$_2$}\xspace}

\newcommand\VulforJ{\textsc{Vul4J}\xspace}
\newcommand\Java{\textsc{Java}\xspace}
\newcommand\JUnit{\textsc{JUnit}\xspace}
\newcommand\vuteco{\textsc{VuTeCo}\xspace}
\newcommand\projectKB{\textsc{ProjectKB}\xspace}
\newcommand\reef{REEF\xspace}
\newcommand\reposvul{\textsc{ReposVul}\xspace}
\newcommand\finding{\textit{Finding}\xspace}
\newcommand\matching{\textit{Matching}\xspace}
\newcommand\finder{\textit{Finder}\xspace}
\newcommand\matcher{\textit{Matcher}\xspace}
\newcommand\codebert{CodeBERT\xspace}
\newcommand\codetfiveplus{CodeT5+\xspace}
\newcommand\unixcoder{UniXcoder\xspace}
\newcommand\codellama{CodeLlama\xspace}
\newcommand\deepseek{DeepSeek Coder\xspace}
\newcommand\qwen{Qwen2.5-Coder\xspace}
\newcommand\codebertFull{\textit{codebert-base}\xspace}
\newcommand\codetfiveplusFull{\textit{codet5p-220m}\xspace}
\newcommand\unixcoderFull{\textit{unixcoder-base}\xspace}
\newcommand\codellamaFull{\textit{CodeLlama-7b-Instruct-hf}\xspace}
\newcommand\deepseekFull{\textit{deepseek-coder-6.7b-instruct}\xspace}
\newcommand\qwenFull{\textit{Qwen2.5-Coder-7B-Instruct}\xspace}
\newcommand\vocab{\textit{Vocab}\xspace}
\newcommand\grepFind{\textit{GrepFind}\xspace}
\newcommand\vocabFind{\textit{VocabFind}\xspace}
\newcommand\vocabFindYake{\textit{VocabFind$_{YAKE}$}\xspace}
\newcommand\vocabFindIden{\textit{VocabFind$_{Iden}$}\xspace}
\newcommand\grepMatch{\textit{GrepMatch}\xspace}
\newcommand\simMatch{\textit{SimMatch}\xspace}
\newcommand\simMatchYake{\textit{SimMatch$_{YAKE}$}\xspace}
\newcommand\simMatchCrm{\textit{SimMatch$_{CRM}$}\xspace}
\newcommand\simMatchCb{\textit{SimMatch$_{CB}$}\xspace}
\newcommand\simMatchCt{\textit{SimMatch$_{CT}$}\xspace}
\newcommand\simMatchUx{\textit{SimMatch$_{UX}$}\xspace}
\newcommand\yakeVocab{\textit{Vocab-YAKE}\xspace}
\newcommand\identVocab{\textit{Vocab-IDEN}\xspace}
\newcommand\simil{\textit{Simil}\xspace}
\newcommand\yakeSimil{\textit{Simil-YAKE}\xspace}
\newcommand\codebertSimil{\textit{Simil-CodeBERT}\xspace}
\newcommand\devNames{\textit{DevNames}\xspace}
\newcommand\fixCommits{\textit{FixCommits}\xspace}
\newcommand\finderPosClass{\textsl{``Security''}\xspace}
\newcommand\finderNegClass{\textsl{``Unclear''}\xspace}
\newcommand\linkerPosClass{\textsl{``Linked''}\xspace}
\newcommand\linkerNegClass{\textsl{``Not-Linked''}\xspace}
\newcommand\matchingPosClass{\textsl{``Matched''}\xspace}
\newcommand\matchingNegClass{\textsl{``Not-Matched''}\xspace}
\newcommand\JavaTransformer{\textsc{JavaTransformer}\xspace}

\newcommand\testforvul{\textsc{Test4Vul}\xspace}

\newcommand\vulforjWitTests{108\xspace}
\newcommand\vulforjNotTests{62,527\xspace}
\newcommand\vulforjNotTestsNonDup{39,542\xspace}
\newcommand\vulforjTestsNonDup{39,650\xspace}
\newcommand\vulforjVulns{79\xspace}
\newcommand\vulforjVulnsWithMetadata{76\xspace}
\newcommand\vulforjProjects{51\xspace}

\newcommand\evalMatchingSimpTrue{105\xspace}
\newcommand\evalMatchingSimpFalse{7,665\xspace}
\newcommand\evalMatchingTrue{105\xspace}
\newcommand\evalMatchingFalse{84,506\xspace}

\newcommand\fzerofive{$F_{0.5}$\xspace}
\newcommand\findingTr{$TR_F$\xspace}
\newcommand\findingTe{$TE_F$\xspace}
\newcommand\findingDe{$DE_F$\xspace}
\newcommand\matchingTr{$TR_M$\xspace}
\newcommand\matchingTe{$TE_M$\xspace}
\newcommand\matchingDe{$DE_M$\xspace}
\newcommand\matchingTrPrime{$TR'_M$\xspace}
\newcommand\matchingTePrime{$TE'_M$\xspace}
\newcommand\matchingDePrime{$DE'_M$\xspace}

\newcommand\vulnsKb{1,992\xspace}
\newcommand\vulnsKbInVulForJ{60\xspace}
\newcommand\vulnsKbNoDescription{19\xspace}
\newcommand\vulnsKbNoProject{61\xspace}
\newcommand\dupRepos{21\xspace}
\newcommand\testRepos{2\xspace}
\newcommand\reposNoTests{388\xspace}
\newcommand\inthewildProjects{427\xspace}
\newcommand\vulnsNoTests{609\xspace}
\newcommand\inthewildVulns{1,238\xspace}
\newcommand\inthewildTests{1,105,491\xspace}
\newcommand\inthewildMatches{5,451,212\xspace}

\newcommand\inthewildFindings{319\xspace}
\newcommand\inthewildFindingsSec{224\xspace}
\newcommand\inthewildFindingsSecProjects{83\xspace}
\newcommand\inthewildFindingsNotSec{92\xspace}
\newcommand\inthewildFindingsInvalid{3\xspace}
\newcommand\inthewildFindingsInvalidL{three\xspace}
\newcommand\inthewildFindingsPrecision{$0.70$\xspace}
\newcommand\inthewildFindingsPrecisionPerc{$70\%$\xspace}
\newcommand\inthewildFindingsAgreement{294\xspace}
\newcommand\inthewildFindingsAgreementPerc{$92\%$\xspace}
\newcommand\inthewildFindingsDisagreement{25\xspace}
\newcommand\inthewildFindingsKappa{$0.81$\xspace}

\newcommand\inthewildMatchings{96\xspace}
\newcommand\inthewildMatchingsCorrect{45\xspace}
\newcommand\inthewildMatchingsIncorrect{51\xspace}
\newcommand\inthewildMatchingsCorrectTests{35\xspace}
\newcommand\inthewildMatchingsCorrectVulns{29\xspace}
\newcommand\inthewildMatchingsCorrectProjects{19\xspace}
\newcommand\inthewildMatchingsPrecision{$0.47$\xspace}
\newcommand\inthewildMatchingsPrecisionPerc{$47\%$\xspace}
\newcommand\inthewildMatchingsAgreement{85\xspace}
\newcommand\inthewildMatchingsAgreementPerc{$89\%$\xspace}
\newcommand\inthewildMatchingsDisagreement{11\xspace}
\newcommand\inthewildMatchingsKappa{$0.77$\xspace}
\newcommand\testforvulSize{259\xspace}

%ACM
\renewcommand{\shortauthors}{Iannone et al.}

\begin{abstract}
Software vulnerabilities are often detected via taint analysis, penetration testing, or fuzzing.
They are also found via \textit{unit tests} that exercise security-sensitive behavior with specific inputs, called \textit{vulnerability-witnessing tests}.
Generative AI models could help developers in writing them, but they require many examples to learn from, which are currently scarce.
This paper introduces \vuteco, an AI-driven framework for collecting examples of vulnerability-witnessing tests from \Java repositories.
\vuteco carries out two tasks: (1) The ``\finding'' task to determine whether a unit test case is security-related, and (2) the ``\matching'' task to relate a test case to the vulnerability it witnesses.
\vuteco addresses the \finding task with \unixcoder, achieving an \fzerofive score of $0.73$ and a precision of $0.83$ on a test set of unit tests from \VulforJ.
The \matching task is addressed using \deepseek, achieving an \fzerofive score of $0.65$ and a precision of $0.75$ on a test set of pairs of unit tests and vulnerabilities from \VulforJ.
\vuteco has been used in the wild on \inthewildProjects \Java projects and \inthewildVulns vulnerabilities, obtaining \inthewildFindingsSec test cases confirmed to be security-related and \inthewildMatchingsCorrectTests tests correctly matched to \inthewildMatchingsCorrectVulns vulnerabilities.
The validated tests were collected in a new dataset called \testforvul.
\vuteco lays the foundation for large-scale retrieval of vulnerability-witnessing tests, enabling future AI models to better understand and generate security unit tests.
\end{abstract}

%ACM

%ACM
\maketitle

\section{Introduction}
\label{sec:intro}
Software vulnerabilities are often detected using techniques such as taint analysis, penetration testing, or fuzzing~\cite{lipp:issta2022,austin:ist2013,Shahriar:csur2012,kaur:coconet2019}, which test a complete application and are usually not integrated into the development workflow.
The ``shift-left'' paradigm encourages a \textit{test-first} approach, in which the test infrastructure to detect vulnerabilities starts at the unit level, akin to how bugs are found~\cite{Gonzalez:esem2021:challenge:sectests}.

\textbf{Security unit tests} trigger vulnerabilities with crafted payloads and use assertions to confirm they are present~\cite{felderer:2016:survey:sectests,Cruzes:2017:agile:sectests,Mohammadi:ase2016:xss:unit}.
Listing~\ref{lst:motivating_jspwiki} shows a unit test for a path traversal vulnerability (CWE-22) affecting \textsc{Apache JSPWiki}, localized in the \texttt{getForwardPage()} method disclosed via CVE-2019-0225.
Lines 3--4 show the official patch for this vulnerability.
If the test method \texttt{testNastyDoPost()} is run on the vulnerable version of \texttt{getForwardPage()}, it fails because the focal method does not return the main wiki page as intended, due to the crafted payload at Line 9.
Hence, the failed assertion at Line 16 confirms the presence of the vulnerability.
We refer to any test case that behaves in this way as a \textbf{vulnerability-witnessing test}~\cite{kang:issta2022:transfer} (a.k.a. a Proof of Vulnerability, PoV~\cite{pinconschi2021comparative, bui:msr2022:vul4j}), or simply \textsl{`witnessing test'} throughout this paper.
These tests differ from ordinary test cases in that they focus on identifying security flaws rather than purely functional defects.

Real-world examples of witnessing tests are collected in \VulforJ~\cite{bui:msr2022:vul4j}, the only \Java dataset containing manually-validated unit tests (\vulforjWitTests in total) matched with \vulforjVulns vulnerabilities affecting \vulforjProjects \Java projects.
The process that led to the creation of \VulforJ mainly consisted of building the projects and running their test. 
This encountered two key problems.
First, building the projects largely failed due to compile errors and missing dependencies, requiring extensive, tentative manual fixes that did not always succeed.
Second, the selected test suites included the tests that existed just after the vulnerability-fixing commit; however, they may not include the witnessing tests, since those could be added in later commits.
Among the 899 vulnerabilities inspected, the authors could only reproduce \vulforjVulns ($\sim$9\%).
Consequently, this strategy for retrieving vulnerability-witnessing tests based on dynamic execution is often unsuccessful (like finding a \textit{needle in a haystack}) and effort-consuming.

Thus, the software security research community is in dire need of more samples of vulnerability witnessing tests.
Some compelling use cases for such data are:
(i) distilling knowledge about the structure of typical tests for each vulnerability type,
(ii) training AI models that can generate or improve security tests~\cite{kang:issta2022:transfer,zhang:2023:llm:sectests},
(iii) verifying the correctness of patches produced by Automated Vulnerability Repair tools~\cite{bui:msr2022:vul4j,liu2019tbar,mohammadi2019automated,gao2021beyond}.
In this paper, we present \textbf{\vuteco} (\underline{\textbf{VU}}lnerability \underline{\textbf{TE}}st \underline{\textbf{CO}}llector), a fully static AI-driven framework that retrieves vulnerability-witnessing tests from \Java test suites.
\vuteco addresses two tasks: (1) the ``\finding'' task to determine whether a test case is security-related, and (2) the ``\matching'' task to pair a test case to the specific vulnerability it is witnessing.

\vuteco has been evaluated in a hold-out set of \VulforJ, where it achieved $0.73$ \fzerofive score and $0.83$ precision in the \finding task.
In the \matching task, \vuteco scored $0.65$ \fzerofive score and $0.75$ precision.
Afterwards, \vuteco was used in a large-scale mining campaign on \textsc{GitHub} (\inthewildProjects projects), where it retrieved \inthewildFindingsSec confirmed security-related test cases, and \inthewildMatchingsCorrectTests tests correctly matched to \inthewildMatchingsCorrectVulns vulnerabilities.
\vuteco provides valuable support for large-scale retrieval of security unit test examples for many downstream applications.

In summary, this paper:
(1) introduces \vuteco, the \textbf{first ever framework} to retrieve vulnerability-witnessing tests in \Java repositories; 
(2) extensively evaluates several AI model types to find the right configuration for \vuteco;
and (3) releases \textbf{\testforvul}, a \textbf{dataset} containing \testforvulSize confirmed security-related unit tests found and matched by \vuteco in the wild.
Data and scripts of this study are released in an online \textbf{replication package}~\cite{appendix} (see Section \hyperref[sec:data:availability]{\textit{``Data Availability''}}).

\begin{listing}[t]
\begin{minted}{java}
// Fix to CVE-2019-0225
public String getForwardPage(HttpServletRequest request) {
|\colorbox{redbg}{\textbf{$-$} return request.getPathInfo();}|
|\colorbox{greenbg}{\textbf{$+$} return ”Wiki.jsp”;}|
}
// Vulnerability-witnessing test
@Test
public void testNastyDoPost() throws Exception {
  MockHttpServletRequest req = new MockHttpServletRequest("/JSPWiki","/wiki/Edit.jsp");
  WikiServlet wikiServlet = new WikiServlet();
  MockServletConfig config = new MockServletConfig();
  config.setServletContext(new MockServletContext("/JSPWiki"));
  wikiServlet.init(config);
  wikiServlet.doPost(req, new MockHttpServletResponse());
  wikiServlet.destroy();
  Assertions.assertEquals("/Wiki.jsp?page=Main&", req.getForwardUrl());
}
\end{minted}
    \caption{Official fix for CVE-2019-0225 and its related witnessing test in \textsc{Apache JSPWiki}.}
    \label{lst:motivating_jspwiki}
\end{listing}

\section{The \vuteco Framework}
\label{sec:vuteco}

\subsection{Overview}
\label{subsec:vuteco_overview}

\vuteco supports two tasks: \finding and \matching.
The \textbf{\finding} task accepts a unit test case, i.e., a \JUnit test method, and tells if it is security-related (\finderPosClass) or if its nature is unclear (\finderNegClass).
The \textbf{\matching} task accepts a unit test case and a textual description (in English) of a known vulnerability and determines if the test case witnesses that vulnerability (\matchingPosClass) or not (\matchingNegClass).
Therefore, both tasks are modeled as \textit{binary classification} problems, which can be run independently from one another (namely, the \matching does not require a prior run of the \finding task).
The two tasks are further detailed in Sections~\ref{subsec:finding} and~\ref{subsec:matching}.

\begin{figure}[t]
    \centering
    \resizebox{.98\linewidth}{!}{
        \includegraphics{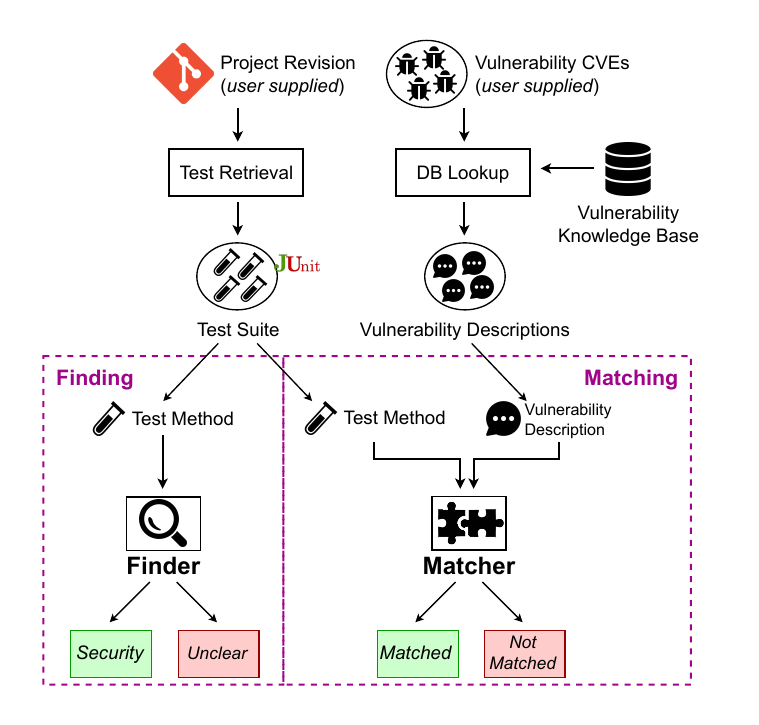}
    }
    \caption{Graphical overview of \vuteco.}
    \label{fig:vuteco-overview}
\end{figure}

Figure~\ref{fig:vuteco-overview} depicts the functioning of \vuteco.
Besides the two AI models, \vuteco also includes a tool that automatically retrieves \JUnit test methods from a given \textsc{Git} repository and vulnerability descriptions from CVE identifiers (see Section~\ref{subsec:retrieval} for details).

\subsection{The \finding Task}
\label{subsec:finding}

\begin{figure}[t]
    \centering
    \resizebox{.98\linewidth}{!}{
        \includegraphics{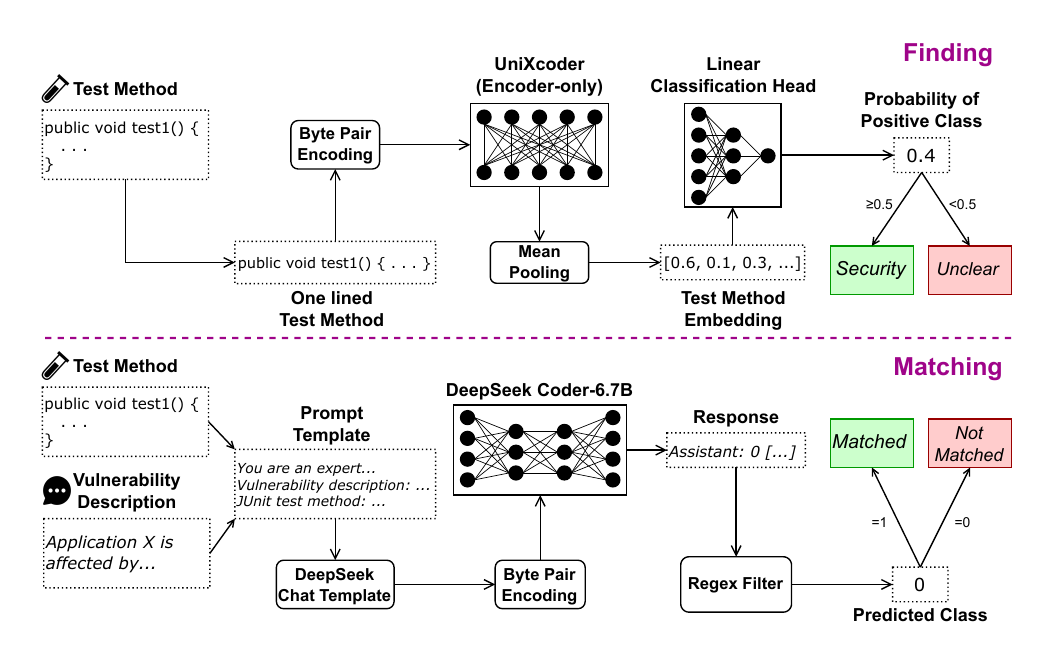}
    }
    \caption{Inner working of \finding and \matching in \vuteco.}
    \label{fig:vuteco-detail}
\end{figure}

The upper part of Figure~\ref{fig:vuteco-detail} illustrates the architecture of the model responsible for the \finding task, which is a neural network based on \unixcoder~\cite{guo2022:unixcoder}.
We hereby refer to this model as \finder.
The choice of \unixcoder resulted from the experimentation designed in Section~\ref{subsec:experiment} and reported in Section~\ref{subsec:experiment-results}.
We selected the checkpoint \unixcoderFull from \textsc{Hugging Face}~\cite{unixcoder:base:hf}, which has been pre-trained on multiple representations of source code, including Abstract Syntax Trees for better understanding.
It experienced six different programming languages, including \Java, as well as natural language from code comments (often English), making it suitable for understanding the content of \JUnit test methods.

Before feeding the input to the model, the test method is stripped of newline characters and consecutive whitespace (including tabs), reducing it to a single line.
The resulting string is tokenized with Byte-Pair Encoding using a vocabulary fitted on CodeSearchNet~\cite{husain2020:codesearchnet}, and then sent to the \unixcoder input layer in ``encoder-only'' mode, i.e., by prepending the \texttt{[CLS]} token and the special token \texttt{[Enc]} to the input (up to 512 encoded tokens).
To obtain the sentence embedding, which represents the entire test method, we perform mean pooling over all token embeddings, as in \unixcoder's official implementation~\cite{unixcoder:github}, yielding a 768-dimensional representation.
On top of this, we added a classification head with two linear layers (with the GELU activation function~\cite{hendrycks2023gaussian:gelu}) of 512 and 128 neurons, respectively, plus an output layer that returns the probability of belonging to the positive class.

The whole model (\unixcoder and the classification head) was trained for ten epochs on a dataset of \finderPosClass and \finderNegClass unit test methods using Weighted Binary Cross-entropy loss to mitigate the large class imbalance.
Additional details on the training setting are reported in Section~\ref{subsec:experiment}.

\subsection{The \matching Task}
\label{subsec:matching}

The lower part of Figure~\ref{fig:vuteco-detail} illustrates the architecture of the model responsible for the \matching task, which uses \deepseek~\cite{guo2024:deepseekcoder}.
We hereby refer to this model as \matcher.
The choice of \deepseek resulted from the experimentation designed in Section~\ref{subsec:experiment} and reported in Section~\ref{subsec:experiment-results}.
\deepseek is a chat-based generative large language model specialized in code-related tasks, including \textit{code understanding}.
We selected the checkpoint 
\deepseekFull from \textsc{Hugging Face}~\cite{qwencoder:instruct:hf}, which has been pre-trained in three sessions:
Two self-supervised sessions (next-token prediction and fill-in-the-middle) to learn to understand individual files, and one supervised session to learn to follow instructions from prompts (i.e., instruction-tuning)~\cite{guo2024:deepseekcoder}.
In total, \deepseek experienced 87 programming languages, including English text from code comments, \textsc{GitHub}  Markdown, and \textsc{StackExchange}.
For all the said reasons, this model can understand the content of \JUnit test methods and vulnerability descriptions.

\vuteco inserts the vulnerability description and the \JUnit test method code (truncated if exceeding the model's maximum length, set to 4,096 tokens) in a natural language prompt as follows:

\prompt{\deepseek Prompt Template}{
\textbf{[System]} You are an expert in unit testing and security testing. Given the following vulnerability description and JUnit test method (it might be truncated if too long), answer with 1 if the test case is likely to identify the described vulnerability in the code under test, or 0 if it is not. Answer with only the number, with no explanation.\\
\textbf{[User]}
Vulnerability Description:\\
\texttt{\{vulnerability\_description\}}\\
JUnit Test Method:\\
\texttt{\{method\_signature\_and\_body\}}\\
\textbf{[Assistant]}
}

The final wording of the system prompt was developed with assistance from ChatGPT (GPT-4o) in May 2025, which helped us refine our initial version (in the replication package~\cite{appendix}) to better suit an LLM.
The placeholders [System], [User], and [Assistant] are not part of the prompt and only indicate the three roles recognized by \deepseek.
This prompt is further transformed by applying the \deepseek's chat template, which starts with the system prompt, then adds `\texttt{\#\#\# Instruction:}' to initiate the user part, and `\texttt{\#\#\# Response:}' to start the assistant part.
The resulting text is tokenized with Byte-Pair Encoding.

At test and inference time, the [Assistant] part is intentionally left empty to induce the model to respond to the system+user prompt, whereas during training, the [Assistant] part contains the ground truth (expected) response---according to the standard practice to fine-tune generative LLMs.
According to the system prompt, the responses are enforced to provide the classification with just 0 or 1.
Hence, the responses are post-processed with a regular expression to extract the first digit encountered (disregarding any additional text), mapping 0 to \matchingNegClass and 1 to \matchingPosClass.

We followed a two-step approach to train the  \deepseek model.
First, we ran a pre-training session for eight epochs on a small dataset of \matchingPosClass and \matchingNegClass pairs involving solely vulnerability-witnessing test methods (i.e., excluding non-security tests), and a fine-tuning session for five epochs on the complete dataset of \matchingPosClass and \matchingNegClass pairs involving non-security tests as well (more details in Section~\ref{subsub:configs}).
Both training steps were modeled as a \textit{Causal Language Modeling} task, computing the Cross Entropy loss only on the [Assistant] part, and employed LoRA (Low-Rank Adaptation) optimization~\cite{hu2021:lora}.
The two training datasets were oversampled with SPAT~\cite{yu:jss2022:spat}, a tool that creates semantically equivalent clones of \Java methods.
We used it to generate further examples of vulnerability-witnessing tests, thereby increasing the number of pairs.
We ran it five times to generate diverse examples (removing any resulting duplicates).
More details on the training setting are reported in Section~\ref{subsec:experiment}.
To ensure that the same pair will always be classified in the same way, \vuteco uses \textit{greedy decoding} during inference, i.e., at each step the model selects the single most probable next token.

\subsection{Tool-assisted Input Retrieval}
\label{subsec:retrieval}

\vuteco is assisted by a \textit{tool} that collects and prepares the input (upper part of Figure~\ref{fig:vuteco-overview}) for the two AI models (the core of \vuteco).
The tool accepts a revision hash (i.e., a commit) of any \textsc{Git}-based repository (remote URL) from the user.
This will be cloned locally, and all its \Java files parsed (using \texttt{javalang}~\cite{javalang}).
Then, it marks as a \textit{``test case''} any \textbf{method} with the following properties:
\begin{enumerate}[leftmargin=*]
    \item it is annotated with \texttt{@Test} (\JUnit 4 and 5) or its class extends \textsc{TestCase} or any subclass of it (for \JUnit 3);
    \item it is not overriding a method defined in class \textsc{TestCase}, like \texttt{run()} or \texttt{getName()} (for \JUnit 3);
    \item it is not a ``lifecycle method'', i.e., annotated with \texttt{@BeforeAll}, \texttt{@AfterAll}, \texttt{@BeforeEach}, or \texttt{@AfterEach};
    \item it returns \texttt{void} if not annotated with \texttt{@TestFactory};
    \item it is not \texttt{abstract}, \texttt{static} or \texttt{private};
    \item its class is not \texttt{abstract}.
\end{enumerate}
Such heuristics were designed based on how the \JUnit guide describes a test case~\cite{junit:guide} and the \textsc{Javadoc} of \JUnit beyond version~3.

Additionally, in the \matching task, the user can supply the tool with a list of CVE identifiers, which are used to look up an internal \textit{knowledge base} of \vulnsKb historical \Java vulnerabilities to fetch their descriptions.
This catalog aggregates and de-duplicates three established datasets of \Java vulnerabilities, i.e., \projectKB~\cite{projectkb:github}, \reef~\cite{wang:ase2023:reef}, and \reposvul~\cite{wang:icse2024:reposvul}.
The catalog also stores additional useful metadata, such as CWE (Common Weakness Enumeration) and known fix commits.
If a CVE identifier is not found in the knowledge base, the tool searches for the missing data on the fly via the \textit{Vulnerability-Lookup} API~\cite{cve:search} and updates the catalog locally.

Afterward, the tool sends the retrieved data to the model responsible for the requested task.
In the \finding case, each collected test case is sent to the \finder model to assess its relevance to security.
In the \matching case, the test cases are paired with the vulnerability descriptions in all possible ways (i.e., the Cartesian product), and each pair is sent to the \matcher model to assess whether the test case witnesses the corresponding vulnerability.
Thus, the ideal use of this tool is to list all CVE identifiers that have ever affected the target project, enabling it to assess every possible match.
\section{Evaluation Design}
\label{sec:design}

We conducted a \textit{two-phase} evaluation.
First, we performed an \textbf{experimental evaluation} to search for the best AI model to carry out the \finding and \matching tasks.
This involved testing several AI model types on a hold-out set originating from \VulforJ---the same source also used for their training.
We involved \textit{six} main models: three based on pre-trained transformers, \codebert~\cite{feng:emnlp2020:codebert}, \codetfiveplus~\cite{wang2023:codet5plus}, and \unixcoder~\cite{guo2022:unixcoder}; and three generative large language models, \codellama~\cite{rozière2024:codellama}, \deepseek~\cite{guo2024:deepseekcoder}, and \qwen~\cite{hui2024:qwen25coder}.
We hereby distinguish the two groups as \textit{code representation models} (CRMs) and \textit{large language models} (LLMs), respectively.
The CRMs transform the input into fixed-length embeddings and process them with a linear classification head to obtain the predicted classes.
The LLMs, instead, are prompted to analyze the input and provide a binary response directly.
We downloaded from \textsc{Hugging Face} the snapshots: \codebertFull, \codetfiveplusFull, \unixcoderFull, \codellamaFull, \deepseekFull, \qwenFull.
This evaluation aimed to assess the suitability of such models for the \finding and \matching tasks using validated data.
For the \finding task, the six models were trained and tested on a collection of test cases extracted from projects in \VulforJ.
In the \matching task, the six models were trained and tested on the same set of test cases, but also paired with vulnerability descriptions that affected the corresponding projects.
Thus, we conducted 12 distinct train-test sessions.

In the second phase, we performed a complementary analysis that examined \vuteco's performance \textbf{in the wild}, i.e., running the best AI models (already mentioned in Section~\ref{sec:vuteco}) resulting from the previous evaluation on a large set of \Java projects outside \VulforJ.
Accordingly, we formulated the following research questions:

\rqbox{1}{What are the \textbf{best AI models} for finding security unit tests and matching them with the right vulnerability?}

\rqbox{2}{How well can \vuteco find security unit tests \textbf{in the wild} and match them with the right vulnerability?}

\subsection[Experimental Evaluation Design (RQ1)]{Experimental Evaluation (RQ$_1$)}
\label{subsec:experiment}

\subsubsection{Data Selection and Preprocessing}
\label{subsubsec:data}

The primary source of data for this experimental evaluation was \VulforJ~\cite{bui:msr2022:vul4j}, as it is the only known source with validated \JUnit test methods matched with the vulnerabilities they witness.
At the time of this paper writing, \VulforJ had \vulforjVulns reproducible vulnerabilities across \vulforjProjects \Java projects.
From these projects, we checked out their patched revisions (i.e., the project versions in which the vulnerability has been fixed and where the witnessing tests have been found) and collected all their test suites using the heuristic described in Section~\ref{subsec:retrieval}.
We fetched metadata---including descriptions---of the \vulforjVulns vulnerabilities appearing in \VulforJ using \vuteco's internal knowledge base (Section \ref{subsec:retrieval}), resulting in \vulforjVulnsWithMetadata vulnerabilities with metadata (three were reported through inaccessible bug trackers rather than CVE).

For the \finding task, we considered all the \vulforjWitTests \JUnit test methods reported \VulforJ as confirmed examples of witnessing tests, labeling them as \finderPosClass (the positive class).
To create the negative class (\finderNegClass), we used all the remaining non-duplicated \JUnit test methods mined from the \vulforjProjects projects in \VulforJ (using the heuristic described in Section~\ref{subsec:retrieval}), totaling \vulforjNotTestsNonDup.
We note that the negative class for the \finding task comprises test cases lacking sufficient evidence of their security relevance, rather than being tests that are entirely unrelated to security (hence the name \finderNegClass).
Then, we split the dataset using stratified sampling (i.e., keeping the same class distribution), creating a training set \findingTr (70\%), a development set \findingDe (10\%), and a test set \findingTe (20\%).

For the \matching task, we created pairs of test cases and the descriptions of the witnessed vulnerabilities in \VulforJ.
For example, the test method \texttt{testSendingStringMessage()} was paired with the vulnerability CVE-2015-0263 (according to the dataset entry VUL4J-3).
Due to the three vulnerabilities discarded previously, the total number of valid pairs was \evalMatchingTrue, which formed the positive class (\matchingPosClass).
To create the negative class (\matchingNegClass), we paired all test methods in the projects in \VulforJ with unrelated vulnerabilities from the same project, producing \evalMatchingFalse invalid pairs.
Akin to the \finding task, we split this dataset using stratified sampling, creating a training set \matchingTr (70\%), a development set \matchingDe (10\%), and a test set \matchingTe (20\%).

Depending on the model type, we prepared the input in different ways.
For the CRMs, the input test methods in both \finding and \matching tasks were flattened into a single line, and any multiple occurrences of whitespace were replaced with a single one---indeed, we observed that removing them could improve performance beyond reducing the input length.
In the \matching task, the description was placed in the test method's \textsc{Javadoc}, exploiting the familiarity that the underlying encoder models have with source code with documentation.
Regarding the LLMs, the prompt for the \matching task was the same as the one presented in Section~\ref{subsec:matching}, while for the \finding task, we used a similar one that omits the vulnerability description.
The difference lies in the central part: \textsl{`Given the following JUnit test method [...] answer with 1 if the test case is likely to identify a vulnerability in the code under test'}.
In both tasks, we did not preprocess the test method code (as done, instead, with the CRMs), but truncated it if it exceeded the model's maximum length (4,096 tokens).

\subsubsection{Performance Indicators}
\label{subsub:eval}

Both the \finding and \matching tasks were modeled as binary classification problems.
Hence, we relied on the traditional metrics to measure the goodness of binary predictions, i.e., \textit{precision} (Pr), \textit{recall} (Re), \textit{F1 score}~\cite{powers:2011:precision:recall,baeza:1999:modern:information:retrieval}.
We also selected the \textit{Matthews's correlation coefficient} (MCC)~\cite{matthews:1975:mcc} as a key indicator due to its reliability in imbalanced problems (like these two), and we reported the absolute number of positive classifications (i.e., True Positives and False Positives) to contextualize all the metrics.

\vuteco's primary goal is to \textbf{maximize the number of correct findings while minimizing incorrect ones}.
The aim is to develop an approach that identifies examples of witnessing tests in software repositories, addressing the challenges outlined in Section~\ref{sec:intro}.
Given this circumstance, we include the \textit{\fzerofive score} in the analysis, which is a varied version of the $F_1$ score where twice as much weight is given to precision compared to recall~\cite{van1979:information:retrieval}.
We prioritized this score in the experimental evaluation because it quantifies the trade-off between precision and recall and aligns with \vuteco's requirements.
Additional metrics, such as the AUC-ROC, are reported in the replication package~\cite{appendix}.

\subsubsection{Model Configuration}
\label{subsub:configs}
We identified key \textbf{factors} that we hypothesized could affect the performance of the AI models.
In Section~\ref{subsec:experiment-results}, we report the best-performing configuration for each AI model (based on the \fzerofive score on the test set).

\textbf{Factor: Data Augmentation.}
In both tasks, we addressed class imbalance in the respective training sets.
We experimented with three mechanisms:
(1) \textbf{\JavaTransformer} (JT)~\cite{rabin:ist2021:javatransformer}, a tool that transforms a \textsc{Java} method (including test methods) into semantically-equivalent clones by applying nine semantic-preserving transformation rules, like renaming variables or adding random logging statements;
(2) \textbf{SPAT}~\cite{yu:jss2022:spat}, another tool creates semantically\-/equivalent clones of \textsc{Java} methods with 18 transformation rules akin to \JavaTransformer;
(3) \textbf{Bootstrapping} (BS), a resampling technique creating exact copies of instances in the minority class (a.k.a. \textit{random oversampling}).
We ran \JavaTransformer (JT) and SPAT five times, as they could generate additional semantically equivalent clones due to random variable names they can synthesize.
For bootstrapping (BS), instead, we set the target imbalance ratio to $0.25$, i.e., re-sampling until the minority instances were 25\% of the total.
We also experimented with a fourth case in which the training data were not augmented.

\textbf{Factor: Loss Function (CRMs only).}
In both tasks, we controlled the loss function for training the CRMs.
We experimented with the standard binary cross-entropy and its \textbf{weighted} version, i.e., where we supplied the weights of the two classes to penalize more the misclassifications made on the minority class (either \finderPosClass or \matchingPosClass)~\cite{Zhang2018:cross:entropy}.
The weights were computed from the training set using the \texttt{compute\_class\_weight()} function of \texttt{scikit-learn}.
Regarding the LLMs, we could only use the standard cross-entropy loss as their training is modeled as a \textit{causal language modeling} task, where the prediction target is a sequence of tokens, rather than one of two possible classes.

\textbf{Factor: Decomposition of \matching (CRMs only).}
In the standard case, training a model for the \matching task consisted of fitting it on \matchingTr.
However, we also modeled this task with a different approach for the three CRMs, i.e., by employing two \textit{sub-models} (of the same architecture, e.g., two \unixcoder models).
The first sub-model aims to determine whether the test method is likely to witness a vulnerability---therefore, it reuses the best configurations resulting from the experimentation made for the \finding task.
The second sub-model implements a ``simplified'' version of the ordinary \matching task by assuming that the input test method is always security-related.
Hence, its three datasets (training, development, and testing) are made by discarding the pairs involving non-witnessing test methods from \matchingTr, \matchingDe, and \matchingTe.
We refer to these altered datasets as \matchingTrPrime, \matchingDePrime, and \matchingTePrime, totaling \evalMatchingSimpTrue \matchingPosClass pairs and \evalMatchingSimpFalse \matchingNegClass pairs.
We separately searched for the optimal configuration for the second sub-model on this simplified \matching task.
We did not employ any decomposition for the LLMs due to the very slow inference time of invoking two models per prediction.

The outputs of the two sub-models are \textbf{integrated} into a single, final judgment to address the ``real'' \matching task.
This integration happened using three approaches.
(1) \textbf{``Meta''} aggregates the logits from both sub-models using a linear layer that outputs a single logit, learning a weighted combination of their predictions.
(2) \textbf{``Fuse''} concatenates the final hidden states (embeddings) from both sub-models, projects them into a 64-dimensional latent space, and then applies another linear layer to output the final logit.
(3) \textbf{``Mask''} returns the logit from the second sub-model only if the first sub-model predicted that the test method is likely to witness a vulnerability ($P$(\finderPosClass)$\ge$$0.5$); otherwise, the logit from the first sub-model is selected as the final output.

\textbf{Factor: Training Mode for \matching.}
Regardless of whether the \matching task is performed by a single model or two sub-models, the availability of the dataset \matchingTrPrime has enabled experimentation with three distinct training modes.
In the \textbf{pre-train} case (PT), the training happens solely on \matchingTrPrime if the \matching task is carried out with one model; otherwise, the two sub-models are just trained separately on \findingTr and \matchingTrPrime, respectively.
\textbf{Fine-tune} (FT) trains the full model directly on \matchingTr, regardless of whether the \matching task is decomposed.
\textbf{Full-train} (PT-FT) employs both \textit{pre-train} and \textit{fine-tune} strategies, sequentially.
These three training modes were tested for all six AI models involved.

\textbf{Hyperparameter Optimization.}
Since the number of factors for the \finding task was fewer than for the \matching task, we could run additional experiments by optimizing some \textit{hyperparameters} to further improve performance, selecting the variants with the highest \fzerofive score and the lowest loss on the development set.
We optimized the ``intensity'' of augmentation mechanisms, i.e., the number of times the \JavaTransformer or SPAT were run during oversampling ($5$ and $15$) and the target imbalance ratio for bootstrapping ($0.25$ and $0.50$).
The size of the two layers of the CRMs' classification head was set to $512$ and $128$; however, for the \finding task, we also optimized their size by searching in the space $\{512, 256\}$ and $\{128, 64\}$, respectively.

\subsubsection{Model Implementation and Training}

%\textbf{Training Setting.}
The models have been implemented with \textsc{PyTorch} and trained with the \textsc{transformers} API.
For all training sessions, the weights were updated using the AdamW optimizer~\cite{loshchilov2019decoupled:adamw}, with a learning rate of $10^{-5}$ for CRMs, and $2\cdot10^{-4}$ for LLMs, both decaying linearly.
The CRMs were trained for 10 epochs, while the LLMs were trained with LoRA (rank 16)~\cite{hu2021:lora} for five epochs.
After each epoch, the model was evaluated on the development set.
Upon completion of training, the checkpoint with the highest \fzerofive score was selected, using the lowest loss as a tiebreaker if necessary.
The classification heads of the CRMs used a dropout probability of $0.1$ between every linear layer to reduce the risk of overfitting~\cite{hinton2012improving:dropout}.

%\textbf{Training Infrastructure.}
The experiment was conducted on a server equipped with an NVIDIA Tesla A100 Core GPU and an Intel Xeon Platinum 8352V CPU.
Counting all combinations of factors (the details are in the replication package~\cite{appendix}), we ran 36 train-testing sessions for the \finding task, of which 24 were for the CRMs and 12 were for the LLMs.
On average, each CRM session took 4 hours, whereas each LLM session took 50 hours.
%(including the time to tune hyperparameters).
For the \matching task, we ran a total of 324 train-testing sessions.
For the CRMs, we had 72 sessions from the case without decomposition into two sub-models, while the other 216 sessions were from the case with two sub-models.
For the LLMs, we had 36 sessions.
On average, each CRM session took 5 hours, whereas each LLM session took 60 hours.
%(including the time to tune hyperparameters).

\subsubsection{Baseline Approaches}
\label{subsub:baselines}

Our study is the first to address the problem of finding security test cases and matching them to vulnerabilities (to the best of our knowledge); thus, there are no direct competitors.
Nevertheless, we developed \textbf{three} \textit{baseline approaches} for the \finding task and \textbf{four} for the \matching task, relying on principles different from those of the CRMs and LLMs.
We experimented with multiple configurations for each baseline.
In Section~\ref{subsec:experiment-results}, we report only the best-performing configuration for each approach (based on the \fzerofive score on the test set).

\textbf{Baselines for \finding.}
The approach \grepFind checks if the test method code contains one or more security-specific keywords, such as \textsl{`secur'}, \textsl{`cve'}, \textsl{`xss'}; if so, it is flagged as \finderPosClass.
We used the set of keywords defined by Zhou and Sharma~\cite{zhou2017automated} and further expanded with additional ones (the full list is in the replication package~\cite{appendix}).
In essence, this approach behaves like the \texttt{grep -e} command.
We conducted experiments varying the required number of matches from $1$ to $5$, testing each condition with and without our extended list, totaling 10 configurations.
The \vocabFind approach, instead, fits a vocabulary of terms on the test methods in the training set \findingTr. Then, it checks if the terms of the input test method adhere to the fitted vocabulary (i.e., the test method shares a similar vocabulary with the known test cases).
If the number of matched terms exceeds the threshold $N$, the input test method is flagged as \finderPosClass.
The terms were extracted in two ways: (i) using YAKE, which picks the $K$ most relevant terms from a text in an unsupervised manner~\cite{campos:2020:yake}, and (ii) retrieving all the identifiers (i.e., variable and method names) in the test case, as they likely indicate the purpose of the test and are not mixed with other irrelevant language-specific keywords.
We refer to these two flavors as \vocabFindYake and \vocabFindIden, respectively.
We experimented with $N$$=$$[1..10]$ (step of $1$) for both flavors.
For \vocabFindYake we experimented with $K$$=$$[5..30]$ (step of $5$), while for \vocabFindIden we experimented with honoring \texttt{camelCase} and \texttt{snake\_case} notations (e.g., if identifiers like \texttt{user\_Password} must be split into \texttt{user} and \texttt{password}), totaling 80 configurations.

\textbf{Baselines for \matching.}
The \grepMatch approach checks if the test method code matches one or more terms appearing in the paired vulnerability description; if so, the pair is flagged as \matchingPosClass.
The terms are vulnerability descriptions obtained via word-level tokenization and removal of English stop words.
We varied the required number of hits from $1$ to $5$.
The \simMatch approach checks the \textit{similarity} between the test method code and the vulnerability description and flags the pair as \linkerPosClass if the similarity score surpasses an arbitrary threshold $T$.
We made two different implementations.
\simMatchYake extracts the keyword sets from both inputs using YAKE~\cite{campos:2020:yake} and compares them using the Jaccard index.
\simMatchCrm uses one of three CRMs, i.e., \codebert~\cite{feng:emnlp2020:codebert}, \codetfiveplus~\cite{wang2023:codet5plus}, and \unixcoder~\cite{unixcoder:base:hf}, to create embeddings for both inputs and compare them using cosine similarity.
For \simMatchYake we experimented with $T$$=$$[0.01..0.05]$ (step of $0.01$) and $K$$=$$[5..30]$ (step of $5$).
For \simMatchCrm we experimented with five similarity thresholds depending on the embedding models: $T$$=$$[0.91..0.95]$ (step of $0.01$) for \codebert, for \codetfiveplus $T$$=$$[0.7..0.9]$ (step of $0.05$), and for \unixcoder $T$$=$$[0.3..0.5]$ (step of $0.05$).
A total of 45 configurations have been evaluated.
Lastly, we developed \fixCommits, which inspects the fix commits of a vulnerability $v$, collects the set of added or modified test methods $TM$, and flags the pairs $(t,v), \forall t \in TM$ as \matchingPosClass.
Any other pair is considered as \matchingNegClass.
This approach relies on the assumption that developers create unit tests alongside the patches to demonstrate that the vulnerability has been successfully fixed.
We note that there is currently insufficient evidence to support that this core assumption always holds, as developers may not create tests when fixing vulnerabilities~\cite{bui:msr2022:vul4j}.
For example, in CVE-2010-0684 of \textsc{ActiveMQ} none of the three fix commits (\texttt{2895197}, \texttt{fed39c3}, and \texttt{9dc43f3}) added any test.
Besides, \fixCommits can only be run when the fix commit is known, which is not the case for all vulnerabilities disclosed on CVE, whereas \vuteco can be used regardless of the fix commit.
Hence, \fixCommits is treated as a heuristic method rather than a ground truth.

\subsection[In-the-wild Evaluation Design (RQ2)]{In-the-wild Evaluation (RQ$_2$)}
\label{subsec:inthewild}

The evaluation in the wild aimed to assess the practical usefulness of \vuteco, i.e., whether it can find new security-related test cases and match them with the right vulnerability.
We prepared \vuteco according to the configuration described in Section~\ref{sec:vuteco}, which resulted from the experimental evaluation (Section~\ref{subsec:experiment}).
This time, we retrained and exported the best models for each task by merging the test sets into their corresponding training sets, since the test sets were no longer needed at this stage.

The dataset comprised open-source \Java projects and their vulnerabilities that we selected from \vuteco's internal knowledge base of \vulnsKb vulnerabilities (introduced in Section~\ref{sec:vuteco}).
First, we discarded \vulnsKbInVulForJ vulnerabilities that were also part of \VulforJ; indeed, they had already been used to train \vuteco, and we already know their witnessing tests.
We discarded \vulnsKbNoDescription more vulnerabilities that had no CVE descriptions, preventing us from running the \matching task.
Then, we assessed the accessibility of their related projects, discarding those without a remote URL or that were no longer accessible, which resulted in the removal of an additional \vulnsKbNoProject vulnerabilities.
We manually reviewed all remaining repositories to remove duplicates (e.g., mirrored repositories hosted outside \textsc{GitHub}) and those containing testing libraries (e.g., \JUnit), since these are not used during deployment.
This step eliminated \dupRepos duplicate repositories and \testRepos testing projects.
Since we had no guidance on when exactly the witnessing tests were added, we selected the latest revision of each repository (i.e., the HEAD of the respective default branches) on May 21, 2025---indeed, even after a vulnerability has been fixed, any witnessing tests that developers wrote might still be part of the test suites.
Therefore, the input given to \vuteco for both tasks was a project's repository (remote URL) and its latest revision.
Besides, for the \matching task, \vuteco also requires a list of historical vulnerabilities that affected it (from the knowledge base).

At this point, we applied the test case retrieval heuristics (Section~\ref{subsec:vuteco_overview}) to ensure the projects had valid test suites.
This revealed \reposNoTests projects without any tests that \vuteco could process.
After all these steps, we obtained \inthewildProjects projects and \inthewildVulns vulnerabilities.
The \inthewildProjects projects had a total of \inthewildTests test cases, which were given to \vuteco for the \finding task. The \finder model required 3 days to classify all test cases.
All valid combinations of test cases and vulnerabilities resulted in \inthewildMatches pairs.
The \matcher model required 36 days to process and classify all of them.

We assessed the correctness of \vuteco in both tasks by measuring its \textit{precision}, i.e., the proportion of positive classifications that are correct.
We extracted all the test cases flagged as \finderPosClass (\finding) and the pairs of test cases and vulnerabilities flagged as \matchingPosClass (\matching).
From this evaluation, we disregarded the \vulforjWitTests vulnerability-witnessing test cases that also appear in \VulforJ---as they have been used to train \vuteco before its deployment in the wild.
We then involved two independent researchers with experience in software security and unit testing.
They were instructed to indicate whether a classification was correct, so that we could compute the true and false positive predictions.
We used Cohen's Kappa score~\cite{cohen:1960:kappa,McHugh:2012:cohen} to measure the agreement between the two inspectors.
The two inspectors discussed all cases of disagreement and reached a consensus to resolve them.
We could not measure recall because it required an infeasible labeling session involving more than one million test methods before launching \vuteco.

\section{Evaluation Results}
\label{sec:results}

\subsection[Experimental Results (RQ1)]{Experimental Results (RQ$_1$)}
\label{subsec:experiment-results}

\subsubsection{\finding Task}
Table~\ref{tab:rq1-fnd} reports the performance of the six evaluated models and the three baseline approaches in their best configurations for the \finding task on \findingTe extracted from \VulforJ (the other configurations are in the replication package~\cite{appendix}).
The best model was \unixcoder~\cite{guo2022:unixcoder}, achieving an \fzerofive score of $0.73$, supported by a high precision of $0.83$.
This means that it can often recognize real vulnerability-witnessing tests, with minimal false positives (just two).
In other words, the tests that it flags are very likely to be security-related.
The good performance of \unixcoder was also confirmed by the high MCC score of $0.61$, indicating a strong positive correlation with the ground truth (following the same interpretation of Pearson's correlation coefficient~\cite{Pearson1895}).
\unixcoder identified 10 out of 21 vulnerability-witnessing tests in \findingTe, resulting in an unremarkable recall score of $0.48$.
Still, this was the highest recall across all tested models.
\unixcoder had slightly lower precision than \qwen, which was $0.88$; however, this was due to a single false positive.

As also reported in Section~\ref{sec:vuteco}, the best configuration of \unixcoder is trained using the Weighted Binary Cross-entropy loss to handle the large class imbalance.
Given this, augmenting the training set was unnecessary (in fact, it was even counterproductive).
Compared to the configuration using the standard Binary Cross-entropy, the improvement in \fzerofive score was 23\%, from $0.56$ to $0.73$.
Thus, we observed that the weighted loss brought the expected benefit.
On the contrary, the data augmentation techniques did not significantly contribute to improving performance---only SPAT~\cite{yu:jss2022:spat} marginally enhanced the \fzerofive score when the standard loss function was used.
The same happened for the runner-up model, i.e., \qwen, whose best configuration (with a $0.66$ \fzerofive score) also adopted the weighted loss and no data augmentation.
From a broader perspective, we observed that CRMs performed comparably to LLMs.
However, given their faster training and inference times, CRMs are the best choice for this task.

\begin{table}[t]
    \centering
    \caption{Performance of the six main AI models and three baselines (best configurations) for the \finding task.}
    \label{tab:rq1-fnd}
    %\small
    \resizebox{0.98\linewidth}{!}{% 
        \begin{tabular}{|c|c|rrrrrrr|}
        \hline
        \rowcolor{black}
        %\textcolor{white}{\textbf{Config}} &
        \multicolumn{2}{|c|}{\textcolor{white}{\textbf{Approach}}}
        & \multicolumn{7}{|c|}{\textcolor{white}{\textbf{Performance}}} \\
        \hline
        %\rowcolor{gray25}
        \multicolumn{2}{|c|}{} &
        \textbf{Pr} &
        \textbf{Re} &
        $\mathbf{F_1}$ &
        $\mathbf{F_{0.5}}$ &
        \textbf{MCC} &
        \textbf{TP} &
        \textbf{FP} \\
        \hline
        % No-aug, WBCE, OneLine, 256-128
        & \codebert & 0.78 & 0.34 & 0.47 & 0.61 & 0.51 & 7 & 2 \\ \cline{2-9}
        % JT-5, BCE, OneLine, 256-128 
        & \codetfiveplus & 0.70 & 0.33 & 0.45 & 0.57 & 0.48 & 7 & 3 \\ \cline{2-9}
        % No-Aug, WBCE, OneLine, 512-128
        %\rowcolor{darkgreen!20}
        \noalign{\vskip +0.3pt}
        \multirow{-3}{*}{\rotatebox{90}{CRM}} & \cellcolor{darkgreen!20} \textbf{\unixcoder} & \cellcolor{darkgreen!20} 0.83 & \cellcolor{darkgreen!20} \textbf{0.48} & \cellcolor{darkgreen!20} \textbf{0.61} & \cellcolor{darkgreen!20} \textbf{0.73} & \cellcolor{darkgreen!20} \textbf{0.63} & \cellcolor{darkgreen!20} \textbf{10} & \cellcolor{darkgreen!20} 2 \\ \hline \Xhline{1.5\arrayrulewidth}
        % Spat-5
        & \codellama & 0.69 & 0.43 & 0.53 & 0.62 & 0.54 & 9 & 4 \\ \cline{2-9}
        % No-Aug
        & \deepseek & 0.69 & 0.43 & 0.53 & 0.62 & 0.54 & 9 & 4 \\ \cline{2-9}
        % No-Aug
        \multirow{-3}{*}{\rotatebox{90}{LLM}} & \qwen & \textbf{0.88} & 0.33 & 0.48 & 0.66 & 0.54 & 7 & \textbf{1} \\
        \hline \Xhline{1.5\arrayrulewidth}
        % 2 matches, not extended
        & \grepFind & 0.01 & 0.24 & 0.03 & 0.02 & 0.05 & 5 & 375 \\ \cline{2-9}
        % 10 matches, 10 keywords
        & \vocabFindYake & 0.08 & 0.05 & 0.06 & 0.07 & 0.06 & 1 & 11 \\ \cline{2-9}
        % 4 matches, no separate, 20 keywords
        \multirow{-3}{*}{\rotatebox{90}{Baseline}} & \vocabFindIden & 0.02 & 0.10 & 0.03 & 0.02 & 0.04 & 2 & 114 \\ \hline
        \end{tabular}%
    }
\end{table}

Among the baseline approaches, none achieved satisfactory results.
\grepFind successfully classified just five witnessing tests in the right class, despite the wide set of keywords that we prepared.
This suggests that developers frequently employ non-obvious terms when writing security-specific test cases.
In any case, this resulted in 375 false positives, rendering it unhelpful.
Even after fitting a vocabulary of terms from the training set using \vocabFindYake and \vocabFindIden, we still experienced poor performance.
To improve these approaches, a more curated list of security-related keywords inferred from real-world examples is required.
However, achieving this requires a larger collection of vulnerability-witnessing test cases, which is actually the primary goal of this work.

\subsubsection{\matching Task}

Table~\ref{tab:rq1-mtc} reports the performance of the six evaluated models and the three baseline approaches in their best configurations for the \matching task on \matchingTe extracted from \VulforJ (the other configurations are in the replication package~\cite{appendix}).
The best model was \deepseek~\cite{guo2024:deepseekcoder}, achieving an \fzerofive score of $0.65$ and a high precision of $0.75$.
This means that it can often validate pairs of test cases and vulnerability descriptions, with minimal false positives (just three).
In other words, the matches that it returns are very likely to be valid.
This is also confirmed by the MCC of $0.57$, indicating a strong positive correlation with the ground truth~\cite{Pearson1895}.

Unlike the \finding task, the best model belongs to the LLM group.
This \deepseek model was trained in ``full-train'' mode (PT-FT), which pre-trained on \matchingTrPrime (the simplified \matching task) for eight epochs and then fine-tuned on \matchingTr (the regular \matching task) for five epochs.
Both training datasets were augmented with SPAT~\cite{yu:jss2022:spat} (ran five times).
This confirmed that the model benefited from the two-phase training.
Indeed, comparing this version with the configuration that only performed fine-tuning (FT) on \matchingTr, the \fzerofive score went from $0.58$ to $0.65$, due to an improved precision (from $0.64$ to $0.75$) and the same recall.
The other LLMs, by contrast, achieved their best performance with only fine-tuning (i.e., pre-training actually lowered their scores).
The effectiveness of two-phase training appears to vary with the specific model, making it an impactful factor.
Furthermore, all LLMs showed consistent benefits from data augmentation.
This is likely due to the inability to employ a weighted loss function as we did with the CRMs.

Regarding the CRMs, the integrated use of two sub-models (one for the \finding part and one for the simplified \matching) was not sufficient to outperform \deepseek.
Nevertheless, the two \unixcoder sub-models ranked second, with perfect precision but very low recall ($0.24$).
The two sub-models have been integrated with the ``Mask'' style (described in Section~\ref{subsub:configs}) and were pre-trained on \findingTr and \matchingTrPrime but without fine-tuning on \matchingTr.
This configuration was significantly better than the equivalent configuration without the integration (i.e., only one \unixcoder), which had an \fzerofive score of $0.01$.
Upon closer inspection of the other CRMs, ``Mask'' was the style that benefited from pre-training alone, as all the other integration styles did not perform well without fine-tuning. This observation aligns with how ``Mask'' operates.
With the exception of the ``Mask'' style, the ``full-train'' mode was the most effective way to train the CRMs.
Thus, if faster inference is needed at the cost of recall (so, with a higher expectation of valid matches misclassified as invalid), \unixcoder is the best option.

\begin{table}[t]
    \centering
    \caption{Performance of the six main AI models and four baselines (best configurations) for the \matching task.}
    \label{tab:rq1-mtc}
    %\small
    \resizebox{0.98\linewidth}{!}{% 
        \begin{tabular}{|c|c|rrrrrrr|}
        \hline
        \rowcolor{black}
        %\textcolor{white}{\textbf{Config}} &
        \multicolumn{2}{|c|}{\textcolor{white}{\textbf{Approach}}}
        & \multicolumn{7}{|c|}{\textcolor{white}{\textbf{Performance}}} \\
        \hline
        %\rowcolor{gray25}
        \multicolumn{2}{|c|}{} &
        \textbf{Pr} &
        \textbf{Re} &
        $\mathbf{F_1}$ &
        $\mathbf{F_{0.5}}$ &
        \textbf{MCC} &
        \textbf{TP} &
        \textbf{FP} \\
        \hline
        % No-integration, PT-FT, BS-25 (PT), BS-50 (FT), BCE, OneLine, 512-128
        & \codebert & 0.60 & 0.29 & 0.39 & 0.49 & 0.41 & 6 & 4 \\ \cline{2-9}
        % FL-Mask, PT, BCE (Finding), JT-5 (Finding), 256-128 (Finding), WBCE (SimplifiedMtc), JT-15 (SimplifiedMtc), 256-64 (SimplifiedMtc), OneLine (both)
        & \codetfiveplus & 0.71 & 0.23 & 0.36 & 0.51 & 0.41 & 5 & 2 \\ \cline{2-9}
        % FL-Mask, PT, WBCE (Finding), No-Aug (Finding), 512-128 (Finding), WBCE (SimplifiedMtc), JT-15 (SimplifiedMtc), 256-64 (SimplifiedMtc), OneLine (both)
        \multirow{-3}{*}{\rotatebox{90}{CRM}} & \unixcoder & \textbf{1.00} & 0.24 & 0.39 & 0.61 & 0.49 & 5 & \textbf{0} \\ \hline \Xhline{1.5\arrayrulewidth}
        % No-integration, FT, SPAT-5
        & \codellama & 0.64 & 0.43 & 0.51 & 0.52 & \textbf{0.58} & 9 & 5 \\ \cline{2-9}
        % No-integration, PT-FT, SPAT-5
        %\rowcolor{darkgreen!20}
        \noalign{\vskip +0.3pt}
        & \cellcolor{darkgreen!20} \textbf{\deepseek} & \cellcolor{darkgreen!20} 0.75 & \cellcolor{darkgreen!20} 0.43 & \cellcolor{darkgreen!20} \textbf{0.55} & \cellcolor{darkgreen!20} \textbf{0.65} & \cellcolor{darkgreen!20} 0.57 & \cellcolor{darkgreen!20} 9 & \cellcolor{darkgreen!20} 3 \\ \cline{2-9}
        % No-integration, FT, BS-25
        \multirow{-3}{*}{\rotatebox{90}{LLM}} & \qwen & 0.86 & 0.29 & 0.43 & 0.61 & 0.50 & 6 & 1 \\
        \hline \Xhline{1.5\arrayrulewidth}
        % 2 Matches
        & \grepMatch & 0.50 & 0.14 & 0.22 & 0.33 & 0.27	& 3 & 3 \\ \cline{2-9}
        % 0.05 Thres, 20 keywords
        & \simMatchYake & 0.01 & 0.10 & 0.02 & 0.01 & 0.03 & 2 & 220 \\ \cline{2-9}
        % UnixCoder 0.45 Thres, 20 keywords
        & \simMatchCrm & 0.33 & 0.05 & 0.09 & 0.15 & 0.12 & 1 & 2 \\ \cline{2-9}
        % 0.94 threshold, 20 keywords
        %\simMatchCb & 0.00 & 1.00 & 0.00 & 0.00 & 0.00 & 21 & 16,800 \\ \hline
        % 0.8, threshold 20 keywords
        %\simMatchCt & 0.00 & 0.10 & 0.01 & 0.00 & 0.01 & 2 & 849 \\ \hline
        % 0.45 Thres, 20 keywords
        %\simMatchUx & 0.33 & 0.05 & 0.09 & 0.15 & 0.12 & 1 & 2 \\ \hline
        %
        \multirow{-4}{*}{\rotatebox{90}{Baseline}} & \fixCommits & 0.30 & \textbf{0.57} & 0.39 & 0.41 & 0.32 & \textbf{12} & 29 \\ \hline
        \end{tabular}%
    }
\end{table}

Overall, the performance in this task is slightly lower than what was observed in the \finding task. This was anticipated because the \matching task appears more challenging: it requires the model to comprehend both the test method and the vulnerability description, which are presented in two different languages (\textsc{Java} and English).

Among the baseline approaches, \simMatch performed poorly.
The \simMatchCrm flavor achieved greater precision, but at the cost of reducing the detection rate almost to zero (only one true positive and two false positives).
Despite its simplicity, \grepMatch (with two required hits) achieved higher precision, i.e., $0.50$.
This method appears suitable as an initial, lightweight approach for quickly matching a few test cases with vulnerabilities (i.e., the ``low-hanging fruit''); however, it also missed many cases ($0.14$ recall).

Overall, the best baseline approach was \fixCommits, achieving $0.41$ \fzerofive score and $0.32$ MCC score.
Although its performance remains lower than that of any of the six AI models, its recall score of $0.57$ was the highest among all---namely, it correctly classified 12 valid pairs, albeit at the cost of several false positives (29).
We shed further light on the unique contribution of this baseline and whether it can complement the findings of the best AI model (\deepseek).
Hence, we selected the valid pairs only (i.e., the 21 positive instances in \matchingTe) and conducted an \textit{overlap analysis} between the two approaches.
We observed that they agreed on five out of 21 cases, with seven cases validated by \fixCommits only, and four by \deepseek. Both missed five matches (false negatives).
In total, their agreement---measured through the Jaccard index---on the positive classifications was $0.31$, meaning that they shared the same judgment in approximately 1 out of 3 cases.
Although the \textit{intersection} of their predicted sets can reduce the number of false positives to zero, and so maximize the precision to maximum, it also drastically reduces the recall to $0.24$ (-$58\%$ compared to \fixCommits).
Hence, in this case, the \textit{union} could be more beneficial as it increases the recall to $0.76$ (+$33\%$ compared to \fixCommits), though this would also reduce precision to $0.5$ (-$33\%$ compared to \deepseek).
In any case, it is worth noting that this joint use is only possible when the fix commits of a vulnerability are known and accessible; otherwise, we are unable to use \fixCommits.

\rqanswer{1}{
\unixcoder is the best model for finding security-related unit tests with minimal false positives, achieving $0.73$ \fzerofive score, $0.83$ precision, and $0.63$ MCC score.
%It was trained using weighted binary cross-entropy loss and without data augmentation.
\deepseek is the best model for matching unit tests and vulnerabilities with minimal false positives, achieving $0.65$ \fzerofive score, $0.75$ precision, and $0.57$ MCC score.
%It has been trained in two phases and benefited from data augmentation with SPAT.
%Precision or recall can be improved if intersected or united with \fixCommits (if fixing commits are known).
}

\subsection[In-the-wild Results (RQ2)]{In-the-wild Results (RQ$_2$)}
\label{subsec:inthewild-results}

\subsubsection{\finding Task}

Based on the results of \rqOne, \vuteco uses \unixcoder for the \finding task.
After processing \inthewildTests test cases (from a total of \inthewildProjects projects), \vuteco flagged \inthewildFindings test methods as \finderPosClass, found in \inthewildFindingsSecProjects projects.
Among these, the manual assessment confirmed that \inthewildFindingsSec covered security-related aspects---i.e., \vuteco scored \inthewildFindingsPrecision precision.
The inspectors agreed on \inthewildFindingsAgreement (\inthewildFindingsAgreementPerc) cases, with \inthewildFindingsKappa Cohen's Kappa score, indicating a \textit{strong} inter-rater agreement.
They jointly reviewed the remaining \inthewildFindingsDisagreement cases with conflicting judgments until consensus was reached.

The precision score in the wild is not much farther from the $0.83$ observed during the experiment in \rqOne---indeed, a non-negligible drop was anticipated.
Thus, the transfer of the \finder model in the wild mostly preserved its capability.
We observed that \vuteco was misled by test cases containing \textit{keywords} commonly associated with vulnerabilities, even though these tests did not actually uncover any security issues.
For instance, a test case in \textsc{Apache Struts} contains terms like \textsl{`bad'}, \textsl{`pollution'}, and \textsl{`inject'}---which are often found in security vocabularies---despite having a different meaning there~\cite{struts:example}.
Besides, we observed that test cases containing literal strings such as \textit{paths} (e.g., URLs or file paths), \textit{shell commands}, \textit{version numbers}, or \textit{hashes} are more likely to be incorrectly flagged.
Such problems could be addressed by introducing fine-grained semantic analyses and improved keyword matching to reduce false positives.

\subsubsection{\matching Task}
Based on the results of \rqOne, \vuteco uses \deepseek for the \matching task.
After processing \inthewildMatches pairs of test cases and vulnerability descriptions, \vuteco flagged \inthewildMatchings pairs as \matchingPosClass, involving \inthewildMatchingsCorrectTests unique tests from \inthewildMatchingsCorrectProjects projects.
Among these, the manual assessment confirmed that \inthewildMatchingsCorrect were correct, i.e., \vuteco scored \inthewildMatchingsPrecision precision.
The inspectors agreed on \inthewildMatchingsAgreement (\inthewildMatchingsAgreementPerc) cases, with \inthewildMatchingsKappa Cohen's Kappa score, indicating a \textit{substantial} inter-rater agreement.
They jointly reviewed the remaining \inthewildMatchingsDisagreement cases with conflicting judgments until consensus was reached.

Unlike the \finding task, here we observed a larger drop in precision from the $0.75$ achieved in \rqOne.
Many of the invalid matches were due to the similarity between the terms in the test case and the CVE description (just as in the \finding case).
This decline in precision can be attributed 
to two main reasons:
(i) the \inthewildProjects projects chosen for this assessment might have no valid match at all, whereas the \vulforjProjects projects used in \rqOne were guaranteed to have at least one valid match;
(ii) the number of vulnerabilities for each project is higher than in \rqOne, particularly for large projects like \textsc{Spring Framework} or \textsc{Jenkins}.
Therefore, there is a natural \textit{distribution shift} that increases the likelihood of false positives.
Despite these issues, the transfer of the \matcher model in the wild was deemed acceptable, though it requires further improvements, such as better prompting and safeguards against misleading terms.

In the end, the \inthewildMatchingsCorrectTests tests correctly matched with the right vulnerability were added to the new dataset \testforvul~\cite{appendix} alongside the \inthewildFindingsSec security-related tests from the \finding task, totaling \testforvulSize.

\rqanswer{2}{
\vuteco found \inthewildFindingsSec security-related test cases in the wild, i.e., \inthewildFindingsPrecisionPerc of all tests returned.
The false positives were primarily due to security-related terms and constant strings appearing in the test code.
\vuteco returned \inthewildMatchingsCorrect correct matches, i.e., \inthewildMatchingsPrecisionPerc of the total matches.
Test code and CVE descriptions with similar vocabulary were matched incorrectly.
}

\section{Discussion}
\label{sec:discussion}

\textbf{The Retrieval of Witnessing Tests.}
The main advantage of \vuteco lies in its fully static nature.
A dynamic assessment would require building an entire project and setting up the environment to execute all test cases, which may not always succeed due to dependency issues.
The extent of the results that \vuteco returned in the wild (\rqTwo) permits manual inspections to be done in a reasonable time.
Indeed, \vuteco greatly reduced the search space by \textit{four orders of magnitude}, from millions to a few hundred cases.
Hence, \vuteco is designed as a lightweight assessment tool to find vulnerability-witnessing tests before proceeding to a more comprehensive dynamic assessment.
Despite the positive results from the experimental evaluation (\rqOne), retrieving vulnerability-witnessing tests remains a \textit{challenging task}.
The main limitation of the AI models we evaluated is their low recall, which did not exceed 0.48 on the \finding task and 0.43 on the \matching task.
This may stem from the training data: \VulforJ contains few, non-diverse security tests, which prevents the models from learning general patterns (for instance, \VulforJ has only six XSS tests and two authentication tests).
Beyond labeling more example tests (also a goal of this paper), the low recall could be mitigated by combining \vuteco's AI models with heuristic methods that achieve higher recall.
Furthermore, we identified the main reasons for \vuteco's false-positive classifications, finding that they may be attributable to the test code's vocabulary.
Tests containing terms and constant strings often related to security led the AI models into error, likely because they were similar to the vocabulary of the positive instances in the training sets for the \finding and \matching tasks.
This problem highlights the need for a \textit{dedicated pre-training session} to adapt the model’s vocabulary to testing and security domains.
Denoising the input test methods to remove excessively long strings and version numbers may also help.
For the \matching task, errors may stem from a \textit{limited understanding of the vulnerability} arising from overly short and vague CVE descriptions. 
This could be mitigated by incorporating additional vulnerability sources, such as issue trackers and mailing lists.

\textbf{The Usefulness of Witnessing Tests.}
The release of \VulforJ in 2022 paved the way for numerous software security tasks.
The ability of \vuteco to find security-related tests helps expand the labeled body of known vulnerability-witnessing tests, enabling activities like the automated generation and improvement of security unit tests~\cite{kang:issta2022:transfer,chen:icse2024:vesta,gao:forge2025:vuleut}.
Nevertheless, the potential applications of vulnerability-witnessing tests extend far beyond this~\cite{pan:icse24:patch:presence}.
For example, the witnessing tests can act as \textit{proofs-of-concept} supporting the automatic generation of realistic exploits~\cite{chen:icse2024:vesta,gao:forge2025:vuleut}.
Besides, they can support the automated vulnerability repair (AVR) process by localizing the vulnerable statements or assessing the correctness of a generated patch~\cite{Mohammadi:issrew2019:unit:tests:repair,Sagodi:ease24:avr:gpt4,Zhou:icse24:vulmaster,Bui:emse2024:apr4vul}.
Researchers can also use such examples to \textit{characterize} security-related tests by distilling the recurring test structure for each vulnerability type.
This would enable comprehending the typical setup actions and assertions involved (further details in the next discussion point).
We also envision using witnessing tests to support the retrieval of vulnerability-contributing commits~\cite{bao2022v,Hinrichs:tosem2025:szz}, as they can be run to triangulate when a vulnerability was introduced in a project.
%; to the best of our knowledge, this task has not yet been investigated.
Additionally, software developers can benefit from access to a variety of example test cases.
For instance, they can reuse tests from past vulnerabilities to address similar issues in their projects.
The dataset \testforvul, which we publicly release for the research community~\cite{appendix}, is designed to achieve such foreseen applications.

\textbf{The Anatomy of Witnessing Tests.}
The retrieval of witnessing tests is challenging, mainly because there is limited empirical knowledge about what such tests look like.
To date, no study has outlined a clear profile of tests that witness vulnerabilities or, more broadly, unit tests focused on security.
The \textit{lack of characterization} of vulnerability tests entails a significant knowledge gap, especially when compared with traditional functional tests.
For instance, we are unaware of the setup required before calling the vulnerable component, what assertions should examine, or the number of tests required to ``cover'' all relevant scenarios for a vulnerability type.
The only commonality between the two test types is that they both aim to identify undesirable behaviors in the code that violate certain requirements or properties.
We believe this lack of knowledge might be ascribable to the difficulty in formulating security requirements at the unit/component level (i.e., methods or classes) since they are often considered at the system level~\cite{Mai:issre18:sec:requirements,felderer:2016:survey:sectests}.
Unfortunately, shedding light on these aspects requires numerous examples of witnessing tests.
In fact, this study was conducted to address this shortage by providing an approach (\vuteco) to expand the knowledge base of witnessing tests and to draw more attention to this topic.
Once a line is drawn between vulnerability-witnessing tests and traditional tests, innovative solutions can be designed to help developers write more security tests.

\section{Threats to Validity}
\label{sec:ttv}

We carefully addressed potential sources of data leakage that could affect the validity of \vuteco's performance.
During the experimental evaluation (\rqOne), the two pre-training datasets for \matching task (see Section~\ref{subsub:configs}), i.e., \findingTr and \matchingTrPrime, were cleaned from test methods and vulnerability descriptions appearing in the test set \matchingTe.
During the evaluation in the wild (\rqTwo), we ignored any known vulnerability-witnessing test case from \VulforJ, as all have been used to train \vuteco's models.
We also acknowledge that the original pre-training of the CRMs and LLMs might have seen some of the test methods used in our evaluation, raising the possibility of data contamination.
Nevertheless, the \finding and \matching classification tasks were different from the objectives of such pre-training, which focused on predicting missing code tokens.
Namely, the models had no prior knowledge about whether a certain test method was security-related or linked to specific vulnerabilities.

The three LLMs experimented with are also available in larger variants, reaching up to $\sim$30 billion parameters.
We opted for the $\sim$7 billion versions due to memory and computational constraints. This provided a reasonable balance between capability and efficiency~\cite{jolicoeurmartineau:2025:tiny}, especially given the large dataset sizes for the two tasks and the many configurations we tested.

The 336 configurations tested in the experimental evaluation (\rqOne) explored key factors affecting performance.
While additional designs were possible, resource limitations made further testing impractical.
\rqOne focused on the model types (CRMs and LLMs), which were hypothesized (and confirmed) to have a relevant impact.

\vuteco has been trained on \JUnit test methods from projects appearing in \VulforJ.
The results cannot be generalized as-is to other programming languages or testing frameworks (e.g., \textsc{TestNG}).
We focused on \Java because of \VulforJ, which provides validated examples of witnessing tests.
\Java remains a relevant language to analyze from a security perspective, as it continues to exhibit new vulnerabilities~\cite{veracode}.
The results could also not be extended to vulnerability types not existing in \Java, e.g., memory-related vulnerabilities, or underrepresented in \VulforJ, such as `OS Command Injection' (CWE-78) that had only one test case~\cite{bui:msr2022:vul4j}.
We partially mitigated this lack of examples during training using data augmentation.

\section{Related Work}
\label{sec:rw}

\textbf{Test Case Classification.}
No research has classified test cases to identify those related to security. Existing studies primarily focused on determining whether a test exhibits \textit{flaky} behavior.
Fatima et al.~\cite{fatima:tse2023:flakify} presented \textsc{Flakify}, a data-driven approach to detect flaky tests in \Java projects.
\textsc{Flakify} leverages a pre-trained CodeBERT and a feed-forward neural network to predict whether a \JUnit test method had a flaky behavior.
\textsc{Flakify} achieved 0.79 and 0.73 F1 scores on two different experiments on \textsc{FlakeFlagger} dataset~\cite{Alshammari:icse2021:FlakeFlagger}, while achieving 0.98 and 0.89 F1 score on \textsc{IDoFT} dataset~\cite{lam:icst2019:idflakies}, outperforming state of the art approaches.
Somewhat similarly, \textsc{FlakyCat} exploits few-shot learning to predict the exact category of flakiness of \JUnit tests~\cite{akli:ast2023:flakycat}.
\textsc{FlakyCat} relies on a pre-trained CodeBERT to create the embeddings of test cases and a Siamese Network to project the embeddings into a space where tests of the same flakiness category appear similar (based on cosine distance).
This is enacted by the Triplet Loss function~\cite{Schroff2015:triplet:loss} during the training.

\textbf{Security Unit Testing.}
Existing works focused on generating test cases for \textit{third-party vulnerabilities}, i.e., those indirectly added through dependencies (e.g., libraries) rather than introduced by the project developers.
Kang et al.~\cite{kang:issta2022:transfer} introduced \textsc{Transfer} to generate security test cases for \Java projects affected by vulnerable library dependencies.
\textsc{Transfer} builds on existing vulnerability-witnessing tests mined from the upstream library project and tries to generate a test case targeting the client project that recreates the same program state generated by the execution of the witnessing test in the original library.
This approach uses a genetic algorithm to find a client test that ``mimics'' the library test's behavior.
\textsc{Transfer} successfully generated security tests for 14 known library vulnerabilities in 23 client projects.
Later, \textsc{Transfer} was extended by Chen et al.~\cite{chen:icse2024:vesta} by including a migration step, which helps ensure that generated tests from client projects are similar to the original vulnerability tests.
Their tool, \textsc{Vesta}, outperforms \textsc{Transfer} by 53\% in test generation effectiveness on a dataset of 30 \Java vulnerabilities.
More recently, Gao et al. introduced \textsc{VulEUT}~\cite{gao:forge2025:vuleut}, which combines static call graph analysis to find out the triggering conditions and a GPT-3.5-Turbo model to generate the concrete unit tests.
\textsc{VulEUT} succeeded in 56/70 cases, whereas VESTA only succeeded in 45.

\section{Conclusion}
\label{sec:conclusion}

We presented \vuteco, a framework for finding security-related tests in \Java projects and matching them to their corresponding vulnerabilities.
The experimental evaluation (\rqOne) identified promising AI models for \finding and \matching tasks, while the assessment in the wild (\rqTwo) demonstrated the usefulness of \vuteco.
The manually-confirmed tests have been collected in a novel dataset, \testforvul, which we have publicly released.
\vuteco is the first solution explicitly tackling the problem of finding security unit tests in software repositories, laying the foundation for future research.

After extensive experimentation, we identified several ways to improve \vuteco.
The maximum input size for \vuteco's AI models could be expanded to include \textit{contextual information}, such as the production code (e.g., the vulnerable class or method) or a natural-language summary of the test cases. 
Namely, the \matching could use additional textual sources like security bug reports or commit messages~\cite{zhou2017automated,le2019automated,bao2022v,nguyen2023multi,iannone:tosem2024:exploit} to improve its accuracy.
Then, the \matching task could replace its ``pair-wise'' classification style with a ``zero-shot'' scheme, allowing it to evaluate a set of candidate vulnerabilities at once rather than individually, thereby reducing the overall number of predictions.
Lastly, \vuteco could include an \textit{automated dynamic assessment} of the retrieved tests to confirm that they can trigger the matched vulnerability as expected.

\section*{Data Availability}
\label{sec:data:availability}
The active version of \vuteco is available at: \textbf{\url{https://github.com/tuhh-softsec/vuteco}}.
\testforvul is available at: \textbf{\url{https://github.com/tuhh-softsec/test4vul}}.
The paper's \textbf{replication package} is available on \textsc{Zenodo}~\cite{appendix} and contains the implementation of \vuteco used in this work, the dataset \testforvul, the scripts to reproduce the experiments, and all raw and processed data.

%ACM
\begin{acks}
\ifthenelse{\boolean{peerreview}}
{Omitted for peer review.}
{
This work was partially supported by EU-funded project Sec4AI4Sec (grant no. 101120393).
}
\end{acks}

\bibliographystyle{ACM-Reference-Format}
\bibliography{biblio}


\begin{thebibliography}{69}

%%% the Web address.

\ifx \showCODEN    \undefined \def \showCODEN     #1{\unskip}     \fi
\ifx \showISBNx    \undefined \def \showISBNx     #1{\unskip}     \fi
\ifx \showISBNxiii \undefined \def \showISBNxiii  #1{\unskip}     \fi
\ifx \showISSN     \undefined \def \showISSN      #1{\unskip}     \fi
\ifx \showLCCN     \undefined \def \showLCCN      #1{\unskip}     \fi
\ifx \shownote     \undefined \def \shownote      #1{#1}          \fi
\ifx \showarticletitle \undefined \def \showarticletitle #1{#1}   \fi
\ifx \showURL      \undefined \def \showURL       {\relax}        \fi
% invisible to TeX
\providecommand\bibfield[2]{#2}
\providecommand\bibinfo[2]{#2}
\providecommand\natexlab[1]{#1}
\providecommand\showeprint[2][]{arXiv:#2}

\bibitem[Akli et~al\mbox{.}(2023)]%
        {akli:ast2023:flakycat}
\bibfield{author}{\bibinfo{person}{Amal Akli}, \bibinfo{person}{Guillaume Haben}, \bibinfo{person}{Sarra Habchi}, \bibinfo{person}{Mike Papadakis}, {and} \bibinfo{person}{Yves Le~Traon}.} \bibinfo{year}{2023}\natexlab{}.
\newblock \showarticletitle{FlakyCat: Predicting Flaky Tests Categories using Few-Shot Learning}. In \bibinfo{booktitle}{\emph{2023 IEEE/ACM International Conference on Automation of Software Test (AST)}}. \bibinfo{pages}{140--151}.
\newblock
\href{https://doi.org/10.1109/AST58925.2023.00018}{doi:\nolinkurl{10.1109/AST58925.2023.00018}}


\bibitem[Alshammari et~al\mbox{.}(2021)]%
        {Alshammari:icse2021:FlakeFlagger}
\bibfield{author}{\bibinfo{person}{Abdulrahman Alshammari}, \bibinfo{person}{Christopher Morris}, \bibinfo{person}{Michael Hilton}, {and} \bibinfo{person}{Jonathan Bell}.} \bibinfo{year}{2021}\natexlab{}.
\newblock \showarticletitle{FlakeFlagger: Predicting Flakiness Without Rerunning Tests}. In \bibinfo{booktitle}{\emph{2021 IEEE/ACM 43rd International Conference on Software Engineering (ICSE)}}. \bibinfo{pages}{1572--1584}.
\newblock
\href{https://doi.org/10.1109/ICSE43902.2021.00140}{doi:\nolinkurl{10.1109/ICSE43902.2021.00140}}


\bibitem[{Apache Flink}(2025)]%
        {struts:example}
\bibfield{author}{\bibinfo{person}{{Apache Flink}}.} \bibinfo{year}{2025}\natexlab{}.
\newblock \bibinfo{title}{{Test case testArrayClassPollutionBlockedByPattern in class ParametersInterceptorTest.java}}.
\newblock \bibinfo{howpublished}{\url{https://github.com/apache/struts/blob/f06bc517ab2bb68521293ddb61fbdae10833085b/core/src/test/java/org/apache/struts2/interceptor/parameter/ParametersInterceptorTest.java\#L286}}.
\newblock
\newblock
\shownote{Online}.


\bibitem[Austin et~al\mbox{.}(2013)]%
        {austin:ist2013}
\bibfield{author}{\bibinfo{person}{Andrew Austin}, \bibinfo{person}{Casper Holmgreen}, {and} \bibinfo{person}{Laurie Williams}.} \bibinfo{year}{2013}\natexlab{}.
\newblock \showarticletitle{A comparison of the efficiency and effectiveness of vulnerability discovery techniques}.
\newblock \bibinfo{journal}{\emph{Information and Software Technology}} \bibinfo{volume}{55}, \bibinfo{number}{7} (\bibinfo{year}{2013}), \bibinfo{pages}{1279--1288}.
\newblock
\showISSN{0950-5849}
\href{https://doi.org/10.1016/j.infsof.2012.11.007}{doi:\nolinkurl{10.1016/j.infsof.2012.11.007}}


\bibitem[Baeza-Yates and Ribeiro-Neto(1999)]%
        {baeza:1999:modern:information:retrieval}
\bibfield{author}{\bibinfo{person}{Ricardo Baeza-Yates} {and} \bibinfo{person}{Berthier Ribeiro-Neto}.} \bibinfo{year}{1999}\natexlab{}.
\newblock \bibinfo{booktitle}{\emph{Modern Information Retrieval}}.
\newblock \bibinfo{publisher}{ACM Press}.
\newblock
\showISBNx{9780201398298}
\showLCCN{99010033}


\bibitem[Bao et~al\mbox{.}(2022)]%
        {bao2022v}
\bibfield{author}{\bibinfo{person}{Lingfeng Bao}, \bibinfo{person}{Xin Xia}, \bibinfo{person}{Ahmed~E Hassan}, {and} \bibinfo{person}{Xiaohu Yang}.} \bibinfo{year}{2022}\natexlab{}.
\newblock \showarticletitle{V-SZZ: automatic identification of version ranges affected by CVE vulnerabilities}. In \bibinfo{booktitle}{\emph{Proceedings of the 44th International Conference on Software Engineering}}. \bibinfo{pages}{2352--2364}.
\newblock


\bibitem[Bui et~al\mbox{.}(2024)]%
        {Bui:emse2024:apr4vul}
\bibfield{author}{\bibinfo{person}{Quang-Cuong Bui}, \bibinfo{person}{Ranindya Paramitha}, \bibinfo{person}{Duc-Ly Vu}, \bibinfo{person}{Fabio Massacci}, {and} \bibinfo{person}{Riccardo Scandariato}.} \bibinfo{year}{2024}\natexlab{}.
\newblock \showarticletitle{APR4Vul: an empirical study of automatic program repair techniques on real-world Java vulnerabilities}.
\newblock \bibinfo{journal}{\emph{Empirical software engineering}} \bibinfo{volume}{29}, \bibinfo{number}{1} (\bibinfo{year}{2024}), \bibinfo{pages}{18}.
\newblock


\bibitem[Bui et~al\mbox{.}(2022)]%
        {bui:msr2022:vul4j}
\bibfield{author}{\bibinfo{person}{Quang-Cuong Bui}, \bibinfo{person}{Riccardo Scandariato}, {and} \bibinfo{person}{Nicol\'{a}s E.~D\'{\i}az Ferreyra}.} \bibinfo{year}{2022}\natexlab{}.
\newblock \showarticletitle{Vul4J: a dataset of reproducible Java vulnerabilities geared towards the study of program repair techniques}. In \bibinfo{booktitle}{\emph{Proceedings of the 19th International Conference on Mining Software Repositories}} (Pittsburgh, Pennsylvania) \emph{(\bibinfo{series}{MSR '22})}. \bibinfo{publisher}{Association for Computing Machinery}, \bibinfo{address}{New York, NY, USA}, \bibinfo{pages}{464–468}.
\newblock
\showISBNx{9781450393034}
\href{https://doi.org/10.1145/3524842.3528482}{doi:\nolinkurl{10.1145/3524842.3528482}}


\bibitem[{c2nes}(2025)]%
        {javalang}
\bibfield{author}{\bibinfo{person}{{c2nes}}.} \bibinfo{year}{2025}\natexlab{}.
\newblock \bibinfo{title}{{javalang}}.
\newblock \bibinfo{howpublished}{\url{https://github.com/c2nes/javalang}}.
\newblock
\newblock
\shownote{Online; accessed 29 July 2025}.


\bibitem[Campos et~al\mbox{.}(2020)]%
        {campos:2020:yake}
\bibfield{author}{\bibinfo{person}{Ricardo Campos}, \bibinfo{person}{Vítor Mangaravite}, \bibinfo{person}{Arian Pasquali}, \bibinfo{person}{Alípio Jorge}, \bibinfo{person}{Célia Nunes}, {and} \bibinfo{person}{Adam Jatowt}.} \bibinfo{year}{2020}\natexlab{}.
\newblock \showarticletitle{YAKE! Keyword extraction from single documents using multiple local features}.
\newblock \bibinfo{journal}{\emph{Information Sciences}}  \bibinfo{volume}{509} (\bibinfo{year}{2020}), \bibinfo{pages}{257--289}.
\newblock
\showISSN{0020-0255}
\href{https://doi.org/10.1016/j.ins.2019.09.013}{doi:\nolinkurl{10.1016/j.ins.2019.09.013}}


\bibitem[Chen et~al\mbox{.}(2024)]%
        {chen:icse2024:vesta}
\bibfield{author}{\bibinfo{person}{Zirui Chen}, \bibinfo{person}{Xing Hu}, \bibinfo{person}{Xin Xia}, \bibinfo{person}{Yi Gao}, \bibinfo{person}{Tongtong Xu}, \bibinfo{person}{David Lo}, {and} \bibinfo{person}{Xiaohu Yang}.} \bibinfo{year}{2024}\natexlab{}.
\newblock \showarticletitle{Exploiting Library Vulnerability via Migration Based Automating Test Generation}. In \bibinfo{booktitle}{\emph{Proceedings of the IEEE/ACM 46th International Conference on Software Engineering}} (<conf-loc>, <city>Lisbon</city>, <country>Portugal</country>, </conf-loc>) \emph{(\bibinfo{series}{ICSE '24})}. \bibinfo{publisher}{Association for Computing Machinery}, \bibinfo{address}{New York, NY, USA}, Article \bibinfo{articleno}{228}, \bibinfo{numpages}{12}~pages.
\newblock
\showISBNx{9798400702174}
\href{https://doi.org/10.1145/3597503.3639583}{doi:\nolinkurl{10.1145/3597503.3639583}}


\bibitem[Cohen(1960)]%
        {cohen:1960:kappa}
\bibfield{author}{\bibinfo{person}{Jacob Cohen}.} \bibinfo{year}{1960}\natexlab{}.
\newblock \showarticletitle{A Coefficient of Agreement for Nominal Scales}.
\newblock \bibinfo{journal}{\emph{Educational and Psychological Measurement}} \bibinfo{volume}{20}, \bibinfo{number}{1} (\bibinfo{year}{1960}), \bibinfo{pages}{37--46}.
\newblock
\href{https://doi.org/10.1177/001316446002000104}{doi:\nolinkurl{10.1177/001316446002000104}}


\bibitem[{Computer Incident Response Center Luxembourg (CIRCL)}(2025)]%
        {cve:search}
\bibfield{author}{\bibinfo{person}{{Computer Incident Response Center Luxembourg (CIRCL)}}.} \bibinfo{year}{2025}\natexlab{}.
\newblock \bibinfo{title}{{Vulnerability-Lookup}}.
\newblock \bibinfo{howpublished}{\url{https://cve.circl.lu/}}.
\newblock
\newblock
\shownote{Online; accessed 29 July 2025}.


\bibitem[Cruzes et~al\mbox{.}(2017)]%
        {Cruzes:2017:agile:sectests}
\bibfield{author}{\bibinfo{person}{Daniela~Soares Cruzes}, \bibinfo{person}{Michael Felderer}, \bibinfo{person}{Tosin~Daniel Oyetoyan}, \bibinfo{person}{Matthias Gander}, {and} \bibinfo{person}{Irdin Pekaric}.} \bibinfo{year}{2017}\natexlab{}.
\newblock \showarticletitle{How is Security Testing Done in Agile Teams? A Cross-Case Analysis of Four Software Teams}. In \bibinfo{booktitle}{\emph{Agile Processes in Software Engineering and Extreme Programming}}. \bibinfo{publisher}{Springer International Publishing}, \bibinfo{address}{Cham}, \bibinfo{pages}{201--216}.
\newblock
\showISBNx{978-3-319-57633-6}


\bibitem[Fatima et~al\mbox{.}(2023)]%
        {fatima:tse2023:flakify}
\bibfield{author}{\bibinfo{person}{Sakina Fatima}, \bibinfo{person}{Taher~A. Ghaleb}, {and} \bibinfo{person}{Lionel Briand}.} \bibinfo{year}{2023}\natexlab{}.
\newblock \showarticletitle{Flakify: A Black-Box, Language Model-Based Predictor for Flaky Tests}.
\newblock \bibinfo{journal}{\emph{IEEE Transactions on Software Engineering}} \bibinfo{volume}{49}, \bibinfo{number}{4} (\bibinfo{year}{2023}), \bibinfo{pages}{1912--1927}.
\newblock
\href{https://doi.org/10.1109/TSE.2022.3201209}{doi:\nolinkurl{10.1109/TSE.2022.3201209}}


\bibitem[Felderer et~al\mbox{.}(2016)]%
        {felderer:2016:survey:sectests}
\bibfield{author}{\bibinfo{person}{Michael Felderer}, \bibinfo{person}{Matthias Büchler}, \bibinfo{person}{Martin Johns}, \bibinfo{person}{Achim~D. Brucker}, \bibinfo{person}{Ruth Breu}, {and} \bibinfo{person}{Alexander Pretschner}.} \bibinfo{year}{2016}\natexlab{}.
\newblock \showarticletitle{Chapter One - Security Testing: A Survey}.
\newblock \bibinfo{series}{Advances in Computers}, Vol.~\bibinfo{volume}{101}. \bibinfo{publisher}{Elsevier}, \bibinfo{pages}{1--51}.
\newblock
\showISSN{0065-2458}
\href{https://doi.org/10.1016/bs.adcom.2015.11.003}{doi:\nolinkurl{10.1016/bs.adcom.2015.11.003}}


\bibitem[Feng et~al\mbox{.}(2020)]%
        {feng:emnlp2020:codebert}
\bibfield{author}{\bibinfo{person}{Zhangyin Feng}, \bibinfo{person}{Daya Guo}, \bibinfo{person}{Duyu Tang}, \bibinfo{person}{Nan Duan}, \bibinfo{person}{Xiaocheng Feng}, \bibinfo{person}{Ming Gong}, \bibinfo{person}{Linjun Shou}, \bibinfo{person}{Bing Qin}, \bibinfo{person}{Ting Liu}, \bibinfo{person}{Daxin Jiang}, {and} \bibinfo{person}{Ming Zhou}.} \bibinfo{year}{2020}\natexlab{}.
\newblock \showarticletitle{{C}ode{BERT}: A Pre-Trained Model for Programming and Natural Languages}. In \bibinfo{booktitle}{\emph{Findings of the Association for Computational Linguistics: EMNLP 2020}}. \bibinfo{publisher}{Association for Computational Linguistics}, \bibinfo{address}{Online}, \bibinfo{pages}{1536--1547}.
\newblock
\href{https://doi.org/10.18653/v1/2020.findings-emnlp.139}{doi:\nolinkurl{10.18653/v1/2020.findings-emnlp.139}}


\bibitem[Gao et~al\mbox{.}(2021)]%
        {gao2021beyond}
\bibfield{author}{\bibinfo{person}{Xiang Gao}, \bibinfo{person}{Bo Wang}, \bibinfo{person}{Gregory~J Duck}, \bibinfo{person}{Ruyi Ji}, \bibinfo{person}{Yingfei Xiong}, {and} \bibinfo{person}{Abhik Roychoudhury}.} \bibinfo{year}{2021}\natexlab{}.
\newblock \showarticletitle{Beyond tests: Program vulnerability repair via crash constraint extraction}.
\newblock \bibinfo{journal}{\emph{ACM Transactions on Software Engineering and Methodology (TOSEM)}} \bibinfo{volume}{30}, \bibinfo{number}{2} (\bibinfo{year}{2021}), \bibinfo{pages}{1--27}.
\newblock


\bibitem[Gao et~al\mbox{.}(2025)]%
        {gao:forge2025:vuleut}
\bibfield{author}{\bibinfo{person}{Yi Gao}, \bibinfo{person}{Xing Hu}, \bibinfo{person}{Zirui Chen}, \bibinfo{person}{Tongtong Xu}, {and} \bibinfo{person}{Xiaohu Yang}.} \bibinfo{year}{2025}\natexlab{}.
\newblock \showarticletitle{Vulnerability-Triggering Test Case Generation from Third-Party Libraries}. In \bibinfo{booktitle}{\emph{2025 IEEE/ACM Second International Conference on AI Foundation Models and Software Engineering (Forge)}}. \bibinfo{pages}{125--135}.
\newblock
\href{https://doi.org/10.1109/Forge66646.2025.00021}{doi:\nolinkurl{10.1109/Forge66646.2025.00021}}


\bibitem[Gonzalez et~al\mbox{.}(2021)]%
        {Gonzalez:esem2021:challenge:sectests}
\bibfield{author}{\bibinfo{person}{Danielle Gonzalez}, \bibinfo{person}{Paola~Peralta Perez}, {and} \bibinfo{person}{Mehdi Mirakhorli}.} \bibinfo{year}{2021}\natexlab{}.
\newblock \showarticletitle{Barriers to Shift-Left Security: The Unique Pain Points of Writing Automated Tests Involving Security Controls}. In \bibinfo{booktitle}{\emph{Proceedings of the 15th ACM / IEEE International Symposium on Empirical Software Engineering and Measurement (ESEM)}} (Bari, Italy) \emph{(\bibinfo{series}{ESEM '21})}. \bibinfo{publisher}{Association for Computing Machinery}, Article \bibinfo{articleno}{11}, \bibinfo{numpages}{12}~pages.
\newblock
\showISBNx{9781450386654}
\href{https://doi.org/10.1145/3475716.3475786}{doi:\nolinkurl{10.1145/3475716.3475786}}


\bibitem[Guo et~al\mbox{.}(2022)]%
        {guo2022:unixcoder}
\bibfield{author}{\bibinfo{person}{Daya Guo}, \bibinfo{person}{Shuai Lu}, \bibinfo{person}{Nan Duan}, \bibinfo{person}{Yanlin Wang}, \bibinfo{person}{Ming Zhou}, {and} \bibinfo{person}{Jian Yin}.} \bibinfo{year}{2022}\natexlab{}.
\newblock \showarticletitle{UniXcoder: Unified Cross-Modal Pre-training for Code Representation}.
\newblock \bibinfo{journal}{\emph{arXiv preprint arXiv:2203.03850}} (\bibinfo{year}{2022}).
\newblock


\bibitem[Guo et~al\mbox{.}(2024)]%
        {guo2024:deepseekcoder}
\bibfield{author}{\bibinfo{person}{Daya Guo}, \bibinfo{person}{Qihao Zhu}, \bibinfo{person}{Dejian Yang}, \bibinfo{person}{Zhenda Xie}, \bibinfo{person}{Kai Dong}, \bibinfo{person}{Wentao Zhang}, \bibinfo{person}{Guanting Chen}, \bibinfo{person}{Xiao Bi}, \bibinfo{person}{Y. Wu}, \bibinfo{person}{Y.~K. Li}, \bibinfo{person}{Fuli Luo}, \bibinfo{person}{Yingfei Xiong}, {and} \bibinfo{person}{Wenfeng Liang}.} \bibinfo{year}{2024}\natexlab{}.
\newblock \bibinfo{title}{DeepSeek-Coder: When the Large Language Model Meets Programming -- The Rise of Code Intelligence}.
\newblock
\showeprint[arxiv]{2401.14196}~[cs.SE]
\urldef\tempurl%
\url{https://arxiv.org/abs/2401.14196}
\showURL{%
\tempurl}


\bibitem[Hendrycks and Gimpel(2023)]%
        {hendrycks2023gaussian:gelu}
\bibfield{author}{\bibinfo{person}{Dan Hendrycks} {and} \bibinfo{person}{Kevin Gimpel}.} \bibinfo{year}{2023}\natexlab{}.
\newblock \showarticletitle{Gaussian Error Linear Units (GELUs)}.
\newblock  (\bibinfo{year}{2023}).
\newblock
\showeprint[arxiv]{1606.08415}~[cs.LG]
\urldef\tempurl%
\url{https://arxiv.org/abs/1606.08415}
\showURL{%
\tempurl}


\bibitem[Hinrichs et~al\mbox{.}(2025)]%
        {Hinrichs:tosem2025:szz}
\bibfield{author}{\bibinfo{person}{Torge Hinrichs}, \bibinfo{person}{Emanuele Iannone}, \bibinfo{person}{Tam\'{a}s Aladics}, \bibinfo{person}{P\'{e}ter Hegedundefineds}, \bibinfo{person}{Andrea De~Lucia}, \bibinfo{person}{Fabio Palomba}, {and} \bibinfo{person}{Riccardo Scandariato}.} \bibinfo{year}{2025}\natexlab{}.
\newblock \showarticletitle{Back to the Roots: Assessing Mining Techniques for Java Vulnerability-Contributing Commits}.
\newblock \bibinfo{journal}{\emph{ACM Trans. Softw. Eng. Methodol.}} (\bibinfo{date}{Sept.} \bibinfo{year}{2025}).
\newblock
\showISSN{1049-331X}
\href{https://doi.org/10.1145/3769105}{doi:\nolinkurl{10.1145/3769105}}
\newblock
\shownote{Just Accepted}.


\bibitem[Hinton et~al\mbox{.}(2012)]%
        {hinton2012improving:dropout}
\bibfield{author}{\bibinfo{person}{Geoffrey~E. Hinton}, \bibinfo{person}{Nitish Srivastava}, \bibinfo{person}{Alex Krizhevsky}, \bibinfo{person}{Ilya Sutskever}, {and} \bibinfo{person}{Ruslan~R. Salakhutdinov}.} \bibinfo{year}{2012}\natexlab{}.
\newblock \showarticletitle{Improving neural networks by preventing co-adaptation of feature detectors}.
\newblock  (\bibinfo{year}{2012}).
\newblock
\showeprint[arxiv]{1207.0580}~[cs.NE]
\urldef\tempurl%
\url{https://arxiv.org/abs/1207.0580}
\showURL{%
\tempurl}


\bibitem[Hu et~al\mbox{.}(2021)]%
        {hu2021:lora}
\bibfield{author}{\bibinfo{person}{Edward~J. Hu}, \bibinfo{person}{Yelong Shen}, \bibinfo{person}{Phillip Wallis}, \bibinfo{person}{Zeyuan Allen-Zhu}, \bibinfo{person}{Yuanzhi Li}, \bibinfo{person}{Shean Wang}, \bibinfo{person}{Lu Wang}, {and} \bibinfo{person}{Weizhu Chen}.} \bibinfo{year}{2021}\natexlab{}.
\newblock \bibinfo{title}{LoRA: Low-Rank Adaptation of Large Language Models}.
\newblock
\showeprint[arxiv]{2106.09685}~[cs.CL]
\urldef\tempurl%
\url{https://arxiv.org/abs/2106.09685}
\showURL{%
\tempurl}


\bibitem[Hui et~al\mbox{.}(2024)]%
        {hui2024:qwen25coder}
\bibfield{author}{\bibinfo{person}{Binyuan Hui}, \bibinfo{person}{Jian Yang}, \bibinfo{person}{Zeyu Cui}, \bibinfo{person}{Jiaxi Yang}, \bibinfo{person}{Dayiheng Liu}, \bibinfo{person}{Lei Zhang}, \bibinfo{person}{Tianyu Liu}, \bibinfo{person}{Jiajun Zhang}, \bibinfo{person}{Bowen Yu}, \bibinfo{person}{Keming Lu}, \bibinfo{person}{Kai Dang}, \bibinfo{person}{Yang Fan}, \bibinfo{person}{Yichang Zhang}, \bibinfo{person}{An Yang}, \bibinfo{person}{Rui Men}, \bibinfo{person}{Fei Huang}, \bibinfo{person}{Bo Zheng}, \bibinfo{person}{Yibo Miao}, \bibinfo{person}{Shanghaoran Quan}, \bibinfo{person}{Yunlong Feng}, \bibinfo{person}{Xingzhang Ren}, \bibinfo{person}{Xuancheng Ren}, \bibinfo{person}{Jingren Zhou}, {and} \bibinfo{person}{Junyang Lin}.} \bibinfo{year}{2024}\natexlab{}.
\newblock \bibinfo{title}{Qwen2.5-Coder Technical Report}.
\newblock
\showeprint[arxiv]{2409.12186}~[cs.CL]
\urldef\tempurl%
\url{https://arxiv.org/abs/2409.12186}
\showURL{%
\tempurl}


\bibitem[Husain et~al\mbox{.}(2020)]%
        {husain2020:codesearchnet}
\bibfield{author}{\bibinfo{person}{Hamel Husain}, \bibinfo{person}{Ho-Hsiang Wu}, \bibinfo{person}{Tiferet Gazit}, \bibinfo{person}{Miltiadis Allamanis}, {and} \bibinfo{person}{Marc Brockschmidt}.} \bibinfo{year}{2020}\natexlab{}.
\newblock \bibinfo{title}{CodeSearchNet Challenge: Evaluating the State of Semantic Code Search}.
\newblock
\showeprint[arxiv]{1909.09436}~[cs.LG]
\urldef\tempurl%
\url{https://arxiv.org/abs/1909.09436}
\showURL{%
\tempurl}


\bibitem[Iannone et~al\mbox{.}(2024)]%
        {iannone:tosem2024:exploit}
\bibfield{author}{\bibinfo{person}{Emanuele Iannone}, \bibinfo{person}{Giulia Sellitto}, \bibinfo{person}{Emanuele Iaccarino}, \bibinfo{person}{Filomena Ferrucci}, \bibinfo{person}{Andrea De~Lucia}, {and} \bibinfo{person}{Fabio Palomba}.} \bibinfo{year}{2024}\natexlab{}.
\newblock \showarticletitle{Early and Realistic Exploitability Prediction of Just-Disclosed Software Vulnerabilities: How Reliable Can It Be?}
\newblock \bibinfo{journal}{\emph{ACM Trans. Softw. Eng. Methodol.}} (\bibinfo{date}{mar} \bibinfo{year}{2024}).
\newblock
\showISSN{1049-331X}
\href{https://doi.org/10.1145/3654443}{doi:\nolinkurl{10.1145/3654443}}
\newblock
\shownote{Just Accepted}.


\bibitem[{Iannone, Emanuele and Bui, Quang-Cuong and Scandariato, Riccardo}(2026)]%
        {appendix}
\bibfield{author}{\bibinfo{person}{{Iannone, Emanuele and Bui, Quang-Cuong and Scandariato, Riccardo}}.} \bibinfo{year}{2026}\natexlab{}.
\newblock \bibinfo{title}{{Paper's Online Appendix}}.
\newblock \bibinfo{howpublished}{\url{https://doi.org/10.5281/zenodo.18258566}}.
\newblock
\newblock
\shownote{Online}.


\bibitem[Jolicoeur-Martineau(2025)]%
        {jolicoeurmartineau:2025:tiny}
\bibfield{author}{\bibinfo{person}{Alexia Jolicoeur-Martineau}.} \bibinfo{year}{2025}\natexlab{}.
\newblock \bibinfo{title}{Less is More: Recursive Reasoning with Tiny Networks}.
\newblock
\showeprint[arxiv]{2510.04871}~[cs.LG]
\urldef\tempurl%
\url{https://arxiv.org/abs/2510.04871}
\showURL{%
\tempurl}


\bibitem[Kang et~al\mbox{.}(2022)]%
        {kang:issta2022:transfer}
\bibfield{author}{\bibinfo{person}{Hong~Jin Kang}, \bibinfo{person}{Truong~Giang Nguyen}, \bibinfo{person}{Bach Le}, \bibinfo{person}{Corina~S. P\u{a}s\u{a}reanu}, {and} \bibinfo{person}{David Lo}.} \bibinfo{year}{2022}\natexlab{}.
\newblock \showarticletitle{Test mimicry to assess the exploitability of library vulnerabilities}. In \bibinfo{booktitle}{\emph{Proceedings of the 31st ACM SIGSOFT International Symposium on Software Testing and Analysis}} \emph{(\bibinfo{series}{ISSTA 2022})}. \bibinfo{publisher}{Association for Computing Machinery}, \bibinfo{address}{New York, NY, USA}, \bibinfo{pages}{276–288}.
\newblock
\showISBNx{9781450393799}
\href{https://doi.org/10.1145/3533767.3534398}{doi:\nolinkurl{10.1145/3533767.3534398}}


\bibitem[Kaur and Nayyar(2020)]%
        {kaur:coconet2019}
\bibfield{author}{\bibinfo{person}{Arvinder Kaur} {and} \bibinfo{person}{Ruchikaa Nayyar}.} \bibinfo{year}{2020}\natexlab{}.
\newblock \showarticletitle{A Comparative Study of Static Code Analysis tools for Vulnerability Detection in C/C++ and JAVA Source Code}.
\newblock \bibinfo{journal}{\emph{Procedia Computer Science}}  \bibinfo{volume}{171} (\bibinfo{year}{2020}), \bibinfo{pages}{2023--2029}.
\newblock
\showISSN{1877-0509}
\href{https://doi.org/10.1016/j.procs.2020.04.217}{doi:\nolinkurl{10.1016/j.procs.2020.04.217}}
\newblock
\shownote{Third International Conference on Computing and Network Communications (CoCoNet'19)}.


\bibitem[Lam et~al\mbox{.}(2019)]%
        {lam:icst2019:idflakies}
\bibfield{author}{\bibinfo{person}{Wing Lam}, \bibinfo{person}{Reed Oei}, \bibinfo{person}{August Shi}, \bibinfo{person}{Darko Marinov}, {and} \bibinfo{person}{Tao Xie}.} \bibinfo{year}{2019}\natexlab{}.
\newblock \showarticletitle{iDFlakies: A Framework for Detecting and Partially Classifying Flaky Tests}. In \bibinfo{booktitle}{\emph{2019 12th IEEE Conference on Software Testing, Validation and Verification (ICST)}}. \bibinfo{pages}{312--322}.
\newblock
\href{https://doi.org/10.1109/ICST.2019.00038}{doi:\nolinkurl{10.1109/ICST.2019.00038}}


\bibitem[Le et~al\mbox{.}(2019)]%
        {le2019automated}
\bibfield{author}{\bibinfo{person}{Triet Huynh~Minh Le}, \bibinfo{person}{Bushra Sabir}, {and} \bibinfo{person}{Muhammad~Ali Babar}.} \bibinfo{year}{2019}\natexlab{}.
\newblock \showarticletitle{Automated software vulnerability assessment with concept drift}. In \bibinfo{booktitle}{\emph{2019 IEEE/ACM 16th International Conference on Mining Software Repositories (MSR)}}. IEEE, \bibinfo{pages}{371--382}.
\newblock


\bibitem[Lipp et~al\mbox{.}(2022)]%
        {lipp:issta2022}
\bibfield{author}{\bibinfo{person}{Stephan Lipp}, \bibinfo{person}{Sebastian Banescu}, {and} \bibinfo{person}{Alexander Pretschner}.} \bibinfo{year}{2022}\natexlab{}.
\newblock \showarticletitle{An empirical study on the effectiveness of static C code analyzers for vulnerability detection}. In \bibinfo{booktitle}{\emph{Proceedings of the 31st ACM SIGSOFT International Symposium on Software Testing and Analysis}} (Virtual, South Korea) \emph{(\bibinfo{series}{ISSTA 2022})}. \bibinfo{publisher}{Association for Computing Machinery}, \bibinfo{address}{New York, NY, USA}, \bibinfo{pages}{544–555}.
\newblock
\showISBNx{9781450393799}
\href{https://doi.org/10.1145/3533767.3534380}{doi:\nolinkurl{10.1145/3533767.3534380}}


\bibitem[Liu et~al\mbox{.}(2019)]%
        {liu2019tbar}
\bibfield{author}{\bibinfo{person}{Kui Liu}, \bibinfo{person}{Anil Koyuncu}, \bibinfo{person}{Dongsun Kim}, {and} \bibinfo{person}{Tegawend{\'e}~F Bissyand{\'e}}.} \bibinfo{year}{2019}\natexlab{}.
\newblock \showarticletitle{TBar: Revisiting template-based automated program repair}. In \bibinfo{booktitle}{\emph{Proceedings of the 28th ACM SIGSOFT international symposium on software testing and analysis}}. \bibinfo{pages}{31--42}.
\newblock


\bibitem[Loshchilov and Hutter(2019)]%
        {loshchilov2019decoupled:adamw}
\bibfield{author}{\bibinfo{person}{Ilya Loshchilov} {and} \bibinfo{person}{Frank Hutter}.} \bibinfo{year}{2019}\natexlab{}.
\newblock \showarticletitle{Decoupled Weight Decay Regularization}.
\newblock  (\bibinfo{year}{2019}).
\newblock
\showeprint[arxiv]{1711.05101}~[cs.LG]
\urldef\tempurl%
\url{https://arxiv.org/abs/1711.05101}
\showURL{%
\tempurl}


\bibitem[Mai et~al\mbox{.}(2018)]%
        {Mai:issre18:sec:requirements}
\bibfield{author}{\bibinfo{person}{Phu~X. Mai}, \bibinfo{person}{Fabrizio Pastore}, \bibinfo{person}{Arda Goknil}, {and} \bibinfo{person}{Lionel~C. Briand}.} \bibinfo{year}{2018}\natexlab{}.
\newblock \showarticletitle{A Natural Language Programming Approach for Requirements-Based Security Testing}. In \bibinfo{booktitle}{\emph{2018 IEEE 29th International Symposium on Software Reliability Engineering (ISSRE)}}. \bibinfo{pages}{58--69}.
\newblock
\href{https://doi.org/10.1109/ISSRE.2018.00017}{doi:\nolinkurl{10.1109/ISSRE.2018.00017}}


\bibitem[Matthews(1975)]%
        {matthews:1975:mcc}
\bibfield{author}{\bibinfo{person}{Brian~W. Matthews}.} \bibinfo{year}{1975}\natexlab{}.
\newblock \showarticletitle{Comparison of the predicted and observed secondary structure of T4 phage lysozyme}.
\newblock \bibinfo{journal}{\emph{Biochimica et Biophysica Acta (BBA) - Protein Structure}} \bibinfo{volume}{405}, \bibinfo{number}{2} (\bibinfo{year}{1975}), \bibinfo{pages}{442--451}.
\newblock


\bibitem[McHugh(2012)]%
        {McHugh:2012:cohen}
\bibfield{author}{\bibinfo{person}{Mary McHugh}.} \bibinfo{year}{2012}\natexlab{}.
\newblock \showarticletitle{Interrater reliability: The kappa statistic}.
\newblock \bibinfo{journal}{\emph{Biochemia medica : \v{c}asopis Hrvatskoga dru\v{s}tva medicinskih biokemi\v{c}ara / HDMB}} \bibinfo{volume}{22}, \bibinfo{number}{3} (\bibinfo{date}{Oct.} \bibinfo{year}{2012}), \bibinfo{pages}{276--82}.
\newblock
\href{https://doi.org/10.11613/bm.2012.031}{doi:\nolinkurl{10.11613/bm.2012.031}}


\bibitem[{Microsoft}(2025a)]%
        {unixcoder:github}
\bibfield{author}{\bibinfo{person}{{Microsoft}}.} \bibinfo{year}{2025}\natexlab{a}.
\newblock \bibinfo{title}{{CodeBERT}}.
\newblock \bibinfo{howpublished}{\url{https://github.com/microsoft/CodeBERT/blob/master/UniXcoder/unixcoder.py\#L80}}.
\newblock
\newblock
\shownote{Online; accessed 29 July 2025}.


\bibitem[{Microsoft}(2025b)]%
        {unixcoder:base:hf}
\bibfield{author}{\bibinfo{person}{{Microsoft}}.} \bibinfo{year}{2025}\natexlab{b}.
\newblock \bibinfo{title}{{UniXcoder-base}}.
\newblock \bibinfo{howpublished}{\url{https://huggingface.co/microsoft/unixcoder-base}}.
\newblock
\newblock
\shownote{Online; accessed 29 July 2025}.


\bibitem[Mohammadi et~al\mbox{.}(2019a)]%
        {mohammadi2019automated}
\bibfield{author}{\bibinfo{person}{Mahmoud Mohammadi}, \bibinfo{person}{Bill Chu}, {and} \bibinfo{person}{Heather~Richter Lipford}.} \bibinfo{year}{2019}\natexlab{a}.
\newblock \showarticletitle{Automated repair of cross-site scripting vulnerabilities through unit testing}. In \bibinfo{booktitle}{\emph{2019 IEEE International symposium on software reliability engineering workshops (ISSREW)}}. IEEE, \bibinfo{pages}{370--377}.
\newblock


\bibitem[Mohammadi et~al\mbox{.}(2016)]%
        {Mohammadi:ase2016:xss:unit}
\bibfield{author}{\bibinfo{person}{Mahmoud Mohammadi}, \bibinfo{person}{Bill Chu}, \bibinfo{person}{Heather~Richter Lipford}, {and} \bibinfo{person}{Emerson Murphy-Hill}.} \bibinfo{year}{2016}\natexlab{}.
\newblock \showarticletitle{Automatic web security unit testing: XSS vulnerability detection}. In \bibinfo{booktitle}{\emph{Proceedings of the 11th International Workshop on Automation of Software Test}} (Austin, Texas) \emph{(\bibinfo{series}{AST '16})}. \bibinfo{publisher}{Association for Computing Machinery}, \bibinfo{address}{New York, NY, USA}, \bibinfo{pages}{78–84}.
\newblock
\showISBNx{9781450341516}
\href{https://doi.org/10.1145/2896921.2896929}{doi:\nolinkurl{10.1145/2896921.2896929}}


\bibitem[Mohammadi et~al\mbox{.}(2019b)]%
        {Mohammadi:issrew2019:unit:tests:repair}
\bibfield{author}{\bibinfo{person}{Mahmoud Mohammadi}, \bibinfo{person}{Bill Chu}, {and} \bibinfo{person}{Heather Richter~Lipford}.} \bibinfo{year}{2019}\natexlab{b}.
\newblock \showarticletitle{Automated Repair of Cross-Site Scripting Vulnerabilities through Unit Testing}. In \bibinfo{booktitle}{\emph{2019 IEEE International Symposium on Software Reliability Engineering Workshops (ISSREW)}}. \bibinfo{pages}{370--377}.
\newblock
\href{https://doi.org/10.1109/ISSREW.2019.00098}{doi:\nolinkurl{10.1109/ISSREW.2019.00098}}


\bibitem[Nguyen et~al\mbox{.}(2023)]%
        {nguyen2023multi}
\bibfield{author}{\bibinfo{person}{Truong~Giang Nguyen}, \bibinfo{person}{Thanh Le-Cong}, \bibinfo{person}{Hong~Jin Kang}, \bibinfo{person}{Ratnadira Widyasari}, \bibinfo{person}{Chengran Yang}, \bibinfo{person}{Zhipeng Zhao}, \bibinfo{person}{Bowen Xu}, \bibinfo{person}{Jiayuan Zhou}, \bibinfo{person}{Xin Xia}, \bibinfo{person}{Ahmed~E Hassan}, {et~al\mbox{.}}} \bibinfo{year}{2023}\natexlab{}.
\newblock \showarticletitle{Multi-granularity detector for vulnerability fixes}.
\newblock \bibinfo{journal}{\emph{IEEE Transactions on Software Engineering}} \bibinfo{volume}{49}, \bibinfo{number}{8} (\bibinfo{year}{2023}), \bibinfo{pages}{4035--4057}.
\newblock


\bibitem[Pan et~al\mbox{.}(2024)]%
        {pan:icse24:patch:presence}
\bibfield{author}{\bibinfo{person}{Zhiyuan Pan}, \bibinfo{person}{Xing Hu}, \bibinfo{person}{Xin Xia}, \bibinfo{person}{Xian Zhan}, \bibinfo{person}{David Lo}, {and} \bibinfo{person}{Xiaohu Yang}.} \bibinfo{year}{2024}\natexlab{}.
\newblock \showarticletitle{PPT4J: Patch Presence Test for Java Binaries}. In \bibinfo{booktitle}{\emph{Proceedings of the IEEE/ACM 46th International Conference on Software Engineering}} (Lisbon, Portugal) \emph{(\bibinfo{series}{ICSE '24})}. \bibinfo{publisher}{Association for Computing Machinery}, \bibinfo{address}{New York, NY, USA}, Article \bibinfo{articleno}{225}, \bibinfo{numpages}{12}~pages.
\newblock
\showISBNx{9798400702174}
\href{https://doi.org/10.1145/3597503.3639231}{doi:\nolinkurl{10.1145/3597503.3639231}}


\bibitem[Pearson and Galton(1895)]%
        {Pearson1895}
\bibfield{author}{\bibinfo{person}{Karl Pearson} {and} \bibinfo{person}{Francis Galton}.} \bibinfo{year}{1895}\natexlab{}.
\newblock \showarticletitle{VII. Note on regression and inheritance in the case of two parents}.
\newblock \bibinfo{journal}{\emph{Proceedings of the Royal Society of London}} \bibinfo{volume}{58}, \bibinfo{number}{347-352} (\bibinfo{year}{1895}), \bibinfo{pages}{240--242}.
\newblock
\showeprint{https://royalsocietypublishing.org/doi/pdf/10.1098/rspl.1895.0041}
\href{https://doi.org/10.1098/rspl.1895.0041}{doi:\nolinkurl{10.1098/rspl.1895.0041}}


\bibitem[Pinconschi et~al\mbox{.}(2021)]%
        {pinconschi2021comparative}
\bibfield{author}{\bibinfo{person}{Eduard Pinconschi}, \bibinfo{person}{Rui Abreu}, {and} \bibinfo{person}{Pedro Ad{\~a}o}.} \bibinfo{year}{2021}\natexlab{}.
\newblock \showarticletitle{A comparative study of automatic program repair techniques for security vulnerabilities}. In \bibinfo{booktitle}{\emph{2021 IEEE 32nd international symposium on software reliability engineering (ISSRE)}}. IEEE, \bibinfo{pages}{196--207}.
\newblock


\bibitem[Powers(2011)]%
        {powers:2011:precision:recall}
\bibfield{author}{\bibinfo{person}{David M.~W. Powers}.} \bibinfo{year}{2011}\natexlab{}.
\newblock \showarticletitle{{Evaluation: From precision, recall and f-measure to roc., informedness, markedness \& correlation}}.
\newblock \bibinfo{journal}{\emph{Journal of Machine Learning Technologies}} \bibinfo{volume}{2}, \bibinfo{number}{1} (\bibinfo{year}{2011}), \bibinfo{pages}{37--63}.
\newblock


\bibitem[{Qwen}(2025)]%
        {qwencoder:instruct:hf}
\bibfield{author}{\bibinfo{person}{{Qwen}}.} \bibinfo{year}{2025}\natexlab{}.
\newblock \bibinfo{title}{{Qwen2.5-Coder-7B}}.
\newblock \bibinfo{howpublished}{\url{https://huggingface.co/Qwen/Qwen2.5-Coder-7B-Instruct}}.
\newblock
\newblock
\shownote{Online; accessed 29 July 2025}.


\bibitem[Rabin et~al\mbox{.}(2021)]%
        {rabin:ist2021:javatransformer}
\bibfield{author}{\bibinfo{person}{Md~Rafiqul~Islam Rabin}, \bibinfo{person}{Nghi~D.Q. Bui}, \bibinfo{person}{Ke Wang}, \bibinfo{person}{Yijun Yu}, \bibinfo{person}{Lingxiao Jiang}, {and} \bibinfo{person}{Mohammad~Amin Alipour}.} \bibinfo{year}{2021}\natexlab{}.
\newblock \showarticletitle{On the generalizability of Neural Program Models with respect to semantic-preserving program transformations}.
\newblock \bibinfo{journal}{\emph{Information and Software Technology}}  \bibinfo{volume}{135} (\bibinfo{year}{2021}), \bibinfo{pages}{106552}.
\newblock
\showISSN{0950-5849}
\href{https://doi.org/10.1016/j.infsof.2021.106552}{doi:\nolinkurl{10.1016/j.infsof.2021.106552}}


\bibitem[Rozière et~al\mbox{.}(2024)]%
        {rozière2024:codellama}
\bibfield{author}{\bibinfo{person}{Baptiste Rozière}, \bibinfo{person}{Jonas Gehring}, \bibinfo{person}{Fabian Gloeckle}, \bibinfo{person}{Sten Sootla}, \bibinfo{person}{Itai Gat}, \bibinfo{person}{Xiaoqing~Ellen Tan}, \bibinfo{person}{Yossi Adi}, \bibinfo{person}{Jingyu Liu}, \bibinfo{person}{Romain Sauvestre}, \bibinfo{person}{Tal Remez}, \bibinfo{person}{Jérémy Rapin}, \bibinfo{person}{Artyom Kozhevnikov}, \bibinfo{person}{Ivan Evtimov}, \bibinfo{person}{Joanna Bitton}, \bibinfo{person}{Manish Bhatt}, \bibinfo{person}{Cristian~Canton Ferrer}, \bibinfo{person}{Aaron Grattafiori}, \bibinfo{person}{Wenhan Xiong}, \bibinfo{person}{Alexandre Défossez}, \bibinfo{person}{Jade Copet}, \bibinfo{person}{Faisal Azhar}, \bibinfo{person}{Hugo Touvron}, \bibinfo{person}{Louis Martin}, \bibinfo{person}{Nicolas Usunier}, \bibinfo{person}{Thomas Scialom}, {and} \bibinfo{person}{Gabriel Synnaeve}.} \bibinfo{year}{2024}\natexlab{}.
\newblock \bibinfo{title}{Code Llama: Open Foundation Models for Code}.
\newblock
\showeprint[arxiv]{2308.12950}~[cs.CL]
\urldef\tempurl%
\url{https://arxiv.org/abs/2308.12950}
\showURL{%
\tempurl}


\bibitem[S\'{a}godi et~al\mbox{.}(2024)]%
        {Sagodi:ease24:avr:gpt4}
\bibfield{author}{\bibinfo{person}{Zolt\'{a}n S\'{a}godi}, \bibinfo{person}{G\'{a}bor Antal}, \bibinfo{person}{Bence Bogenf\"{u}rst}, \bibinfo{person}{Martin Isztin}, \bibinfo{person}{P\'{e}ter Hegedundefineds}, {and} \bibinfo{person}{Rudolf Ferenc}.} \bibinfo{year}{2024}\natexlab{}.
\newblock \showarticletitle{Reality Check: Assessing GPT-4 in Fixing Real-World Software Vulnerabilities}. In \bibinfo{booktitle}{\emph{Proceedings of the 28th International Conference on Evaluation and Assessment in Software Engineering}} (Salerno, Italy) \emph{(\bibinfo{series}{EASE '24})}. \bibinfo{publisher}{Association for Computing Machinery}, \bibinfo{address}{New York, NY, USA}, \bibinfo{pages}{252–261}.
\newblock
\showISBNx{9798400717017}
\href{https://doi.org/10.1145/3661167.3661207}{doi:\nolinkurl{10.1145/3661167.3661207}}


\bibitem[{SAP}(2025)]%
        {projectkb:github}
\bibfield{author}{\bibinfo{person}{{SAP}}.} \bibinfo{year}{2025}\natexlab{}.
\newblock \bibinfo{title}{{project-kb}}.
\newblock \bibinfo{howpublished}{\url{https://github.com/SAP/project-kb}}.
\newblock
\newblock
\shownote{Online; accessed 29 July 2025}.


\bibitem[Schroff et~al\mbox{.}(2015)]%
        {Schroff2015:triplet:loss}
\bibfield{author}{\bibinfo{person}{Florian Schroff}, \bibinfo{person}{Dmitry Kalenichenko}, {and} \bibinfo{person}{James Philbin}.} \bibinfo{year}{2015}\natexlab{}.
\newblock \showarticletitle{FaceNet: A unified embedding for face recognition and clustering}. In \bibinfo{booktitle}{\emph{2015 IEEE Conference on Computer Vision and Pattern Recognition (CVPR)}}. \bibinfo{publisher}{IEEE}.
\newblock
\href{https://doi.org/10.1109/cvpr.2015.7298682}{doi:\nolinkurl{10.1109/cvpr.2015.7298682}}


\bibitem[Shahriar and Zulkernine(2012)]%
        {Shahriar:csur2012}
\bibfield{author}{\bibinfo{person}{Hossain Shahriar} {and} \bibinfo{person}{Mohammad Zulkernine}.} \bibinfo{year}{2012}\natexlab{}.
\newblock \showarticletitle{Mitigating program security vulnerabilities: Approaches and challenges}.
\newblock \bibinfo{journal}{\emph{ACM Comput. Surv.}} \bibinfo{volume}{44}, \bibinfo{number}{3}, Article \bibinfo{articleno}{11} (\bibinfo{date}{jun} \bibinfo{year}{2012}), \bibinfo{numpages}{46}~pages.
\newblock
\showISSN{0360-0300}
\href{https://doi.org/10.1145/2187671.2187673}{doi:\nolinkurl{10.1145/2187671.2187673}}


\bibitem[{Stefan Bechtold, Sam Brannen, Johannes Link, Matthias Merdes, Marc Philipp, Juliette de Rancourt, Christian Stein}(2025)]%
        {junit:guide}
\bibfield{author}{\bibinfo{person}{{Stefan Bechtold, Sam Brannen, Johannes Link, Matthias Merdes, Marc Philipp, Juliette de Rancourt, Christian Stein}}.} \bibinfo{year}{2025}\natexlab{}.
\newblock \bibinfo{title}{{JUnit 5 User Guide}}.
\newblock \bibinfo{howpublished}{\url{https://docs.junit.org/current/user-guide/}}.
\newblock
\newblock
\shownote{Online; accessed 29 July 2025}.


\bibitem[Van~Rijsbergen(1979)]%
        {van1979:information:retrieval}
\bibfield{author}{\bibinfo{person}{C.J. Van~Rijsbergen}.} \bibinfo{year}{1979}\natexlab{}.
\newblock \bibinfo{booktitle}{\emph{Information Retrieval}}.
\newblock \bibinfo{publisher}{Butterworths}.
\newblock
\showISBNx{9780408709293}
\showLCCN{78040725}
\urldef\tempurl%
\url{https://books.google.de/books?id=t-pTAAAAMAAJ}
\showURL{%
\tempurl}


\bibitem[Veracode(2024)]%
        {veracode}
\bibfield{author}{\bibinfo{person}{Veracode}.} \bibinfo{year}{2024}\natexlab{}.
\newblock \bibinfo{title}{{State of Software Security}}.
\newblock \bibinfo{howpublished}{\url{https://www.veracode.com/wp-content/uploads/2024/06/SOSS-Report-2024.pdf}}.
\newblock
\newblock
\shownote{Online; accessed 29 July 2025}.


\bibitem[Wang et~al\mbox{.}(2024b)]%
        {wang:ase2023:reef}
\bibfield{author}{\bibinfo{person}{Chaozheng Wang}, \bibinfo{person}{Zongjie Li}, \bibinfo{person}{Yun Peng}, \bibinfo{person}{Shuzheng Gao}, \bibinfo{person}{Sirong Chen}, \bibinfo{person}{Shuai Wang}, \bibinfo{person}{Cuiyun Gao}, {and} \bibinfo{person}{Michael~R. Lyu}.} \bibinfo{year}{2024}\natexlab{b}.
\newblock \showarticletitle{REEF: A Framework for Collecting Real-World Vulnerabilities and Fixes}. In \bibinfo{booktitle}{\emph{Proceedings of the 38th IEEE/ACM International Conference on Automated Software Engineering}} (Echternach, Luxembourg) \emph{(\bibinfo{series}{ASE '23})}. \bibinfo{publisher}{IEEE Press}, \bibinfo{pages}{1952–1962}.
\newblock
\showISBNx{9798350329964}
\href{https://doi.org/10.1109/ASE56229.2023.00199}{doi:\nolinkurl{10.1109/ASE56229.2023.00199}}


\bibitem[Wang et~al\mbox{.}(2024a)]%
        {wang:icse2024:reposvul}
\bibfield{author}{\bibinfo{person}{Xinchen Wang}, \bibinfo{person}{Ruida Hu}, \bibinfo{person}{Cuiyun Gao}, \bibinfo{person}{Xin-Cheng Wen}, \bibinfo{person}{Yujia Chen}, {and} \bibinfo{person}{Qing Liao}.} \bibinfo{year}{2024}\natexlab{a}.
\newblock \showarticletitle{ReposVul: A Repository-Level High-Quality Vulnerability Dataset}. In \bibinfo{booktitle}{\emph{Proceedings of the 2024 IEEE/ACM 46th International Conference on Software Engineering: Companion Proceedings}} (Lisbon, Portugal) \emph{(\bibinfo{series}{ICSE-Companion '24})}. \bibinfo{publisher}{Association for Computing Machinery}, \bibinfo{address}{New York, NY, USA}, \bibinfo{pages}{472–483}.
\newblock
\showISBNx{9798400705021}
\href{https://doi.org/10.1145/3639478.3647634}{doi:\nolinkurl{10.1145/3639478.3647634}}


\bibitem[Wang et~al\mbox{.}(2023)]%
        {wang2023:codet5plus}
\bibfield{author}{\bibinfo{person}{Yue Wang}, \bibinfo{person}{Hung Le}, \bibinfo{person}{Akhilesh~Deepak Gotmare}, \bibinfo{person}{Nghi~D.Q. Bui}, \bibinfo{person}{Junnan Li}, {and} \bibinfo{person}{Steven C.~H. Hoi}.} \bibinfo{year}{2023}\natexlab{}.
\newblock \showarticletitle{CodeT5+: Open Code Large Language Models for Code Understanding and Generation}.
\newblock \bibinfo{journal}{\emph{arXiv preprint}} (\bibinfo{year}{2023}).
\newblock


\bibitem[Yu et~al\mbox{.}(2022)]%
        {yu:jss2022:spat}
\bibfield{author}{\bibinfo{person}{Shiwen Yu}, \bibinfo{person}{Ting Wang}, {and} \bibinfo{person}{Ji Wang}.} \bibinfo{year}{2022}\natexlab{}.
\newblock \showarticletitle{Data Augmentation by Program Transformation}.
\newblock \bibinfo{journal}{\emph{Journal of Systems and Software}}  \bibinfo{volume}{190} (\bibinfo{year}{2022}), \bibinfo{pages}{111304}.
\newblock
\showISSN{0164-1212}
\href{https://doi.org/10.1016/j.jss.2022.111304}{doi:\nolinkurl{10.1016/j.jss.2022.111304}}


\bibitem[Zhang et~al\mbox{.}(2023)]%
        {zhang:2023:llm:sectests}
\bibfield{author}{\bibinfo{person}{Ying Zhang}, \bibinfo{person}{Wenjia Song}, \bibinfo{person}{Zhengjie Ji}, \bibinfo{person}{Danfeng}, \bibinfo{person}{Yao}, {and} \bibinfo{person}{Na Meng}.} \bibinfo{year}{2023}\natexlab{}.
\newblock \showarticletitle{How well does LLM generate security tests?}
\newblock  (\bibinfo{year}{2023}).
\newblock
\showeprint[arxiv]{2310.00710}
\urldef\tempurl%
\url{https://arxiv.org/abs/2310.00710}
\showURL{%
\tempurl}


\bibitem[Zhang and Sabuncu(2018)]%
        {Zhang2018:cross:entropy}
\bibfield{author}{\bibinfo{person}{Zhilu Zhang} {and} \bibinfo{person}{Mert~R. Sabuncu}.} \bibinfo{year}{2018}\natexlab{}.
\newblock \showarticletitle{Generalized cross entropy loss for training deep neural networks with noisy labels}. In \bibinfo{booktitle}{\emph{Proceedings of the 32nd International Conference on Neural Information Processing Systems}} (Montr\'{e}al, Canada) \emph{(\bibinfo{series}{NIPS'18})}. \bibinfo{publisher}{Curran Associates Inc.}, \bibinfo{address}{Red Hook, NY, USA}, \bibinfo{pages}{8792–8802}.
\newblock


\bibitem[Zhou et~al\mbox{.}(2024)]%
        {Zhou:icse24:vulmaster}
\bibfield{author}{\bibinfo{person}{Xin Zhou}, \bibinfo{person}{Kisub Kim}, \bibinfo{person}{Bowen Xu}, \bibinfo{person}{Donggyun Han}, {and} \bibinfo{person}{David Lo}.} \bibinfo{year}{2024}\natexlab{}.
\newblock \showarticletitle{Out of Sight, Out of Mind: Better Automatic Vulnerability Repair by Broadening Input Ranges and Sources}. In \bibinfo{booktitle}{\emph{Proceedings of the IEEE/ACM 46th International Conference on Software Engineering}} (Lisbon, Portugal) \emph{(\bibinfo{series}{ICSE '24})}. \bibinfo{publisher}{Association for Computing Machinery}, \bibinfo{address}{New York, NY, USA}, Article \bibinfo{articleno}{88}, \bibinfo{numpages}{13}~pages.
\newblock
\showISBNx{9798400702174}
\href{https://doi.org/10.1145/3597503.3639222}{doi:\nolinkurl{10.1145/3597503.3639222}}


\bibitem[Zhou and Sharma(2017)]%
        {zhou2017automated}
\bibfield{author}{\bibinfo{person}{Yaqin Zhou} {and} \bibinfo{person}{Asankhaya Sharma}.} \bibinfo{year}{2017}\natexlab{}.
\newblock \showarticletitle{Automated identification of security issues from commit messages and bug reports}. In \bibinfo{booktitle}{\emph{Proceedings of the 2017 11th joint meeting on foundations of software engineering}}. \bibinfo{pages}{914--919}.
\newblock


\end{thebibliography}

\end{document}